\documentclass[fleqn,10pt]{wlscirep}
\usepackage{amsmath}
\usepackage{amssymb}
\usepackage{amsfonts}
\usepackage[utf8]{inputenc}
\usepackage[T1]{fontenc}
\usepackage{bm}
\usepackage{siunitx}
\usepackage{soul}
\usepackage[normalem]{ulem}
\usepackage[skins,theorems]{tcolorbox}
\tcbset{highlight math style={enhanced,
  colframe=red,colback=white,arc=0pt,boxrule=1pt}}
\usepackage{hyperref}
\usepackage{lineno}
\usepackage{ulem}
\hypersetup{
     colorlinks=true,
     linkcolor=blue,
     filecolor=blue,
     citecolor = blue,      
     urlcolor=blue
     }
\usepackage{tcolorbox}
\usepackage{tabularx}

\usepackage{array}
\usepackage{colortbl,ulem}
\tcbuselibrary{skins}

\newcommand\redout{\bgroup\markoverwith
{\textcolor{black}{\rule[.5ex]{2pt}{0.4pt}}}\ULon}

\newcolumntype{Y}{>{\raggedleft\arraybackslash}X}

\tcbset{tab1/.style={fonttitle=\bfseries\large,fontupper=\normalsize\sffamily,
colback=yellow!10!white,colframe=red!75!black,colbacktitle=yellow!40!white,
coltitle=black,center title,freelance,frame code={
\foreach \n in {north east,north west,south east,south west}
{\path [fill=red!75!black] (interior.\n) circle (3mm); };},}}

\tcbset{tab2/.style={enhanced,fonttitle=\bfseries,fontupper=\normalsize\sffamily,
colback=yellow!10!white,colframe=red!50!black,colbacktitle=yellow!40!white,
coltitle=black,center title}}

\title{Glass-like anomalies and unconventional thermoelectric transport in chimney ladder crystals}

\author[1,$*$]{Srinivas V. Mandyam}
\author[2,3,$*$]{Weicen Dong}
\author[4,5,$*$]{Xiaoxian Yan}
\author[2,4,$*$]{Binru Zhao}
\author[6]{Yasong Wu}
\author[6,7]{Chunhao Su}
\author[8]{Elen Duverger-N\'edellec}
\author[9]{Junfa Lin}
\author[9,10,11]{Tianlong Xia}
\author[12]{Zhiying Zhao}
\author[13]{Xi Chen}
\author[6,$\dagger$]{Jiong Yang}
\author[4,14,$\dagger$]{Jie Ma}
\author[4,5,$\dagger$]{Hui Xing}
\author[1,$\dagger$]{F. Malte Grosche}
\author[2,3,$\dagger$]{Matteo Baggioli}

\affil[1]{Cavendish Laboratory, University of Cambridge, JJ Thomson Avenue, CB30HE Cambridge, U.K.}
\affil[2]{Wilczek Quantum Center, School of Physics and Astronomy, Shanghai Jiao Tong University, Shanghai 200240, China}
\affil[3]{Shanghai Research Center for Quantum Sciences, Shanghai 201315,China}
\affil[4]{Key Laboratory of Artificial Structures and Quantum Control, School of Physics and Astronomy, Shanghai Jiao Tong University, Shanghai, 200240, China}
\affil[5]{Shanghai Center for Complex Physics, School of Physics and Astronomy, Shanghai Jiao Tong University, Shanghai, China}
\affil[6]{Materials Genome Institute, State Key Laboratory of Advanced Refractories, Shanghai University, Shanghai 200444, China}
\affil[7]{Qianweichang College, Shanghai University, Shanghai 200444, China}
\affil[8]{University of Bordeaux, CNRS, Bordeaux INP, ICMCB, UMR 5026, F-33600 Pessac, France}
\affil[9]{School of Physics and Beijing Key Laboratory of Opto-electronic Functional Materials $\&$ Micro-nano Devices, Renmin University of China, Beijing 100872, China}
\affil[10]{Key Laboratory of Quantum State Construction and Manipulation (Ministry of Education), Renmin University of China, Beijing, 100872, China}
\affil[11]{Laboratory for Neutron Scattering, Renmin University of China, Beijing 100872, China}
\affil[12]{State Key Laboratory of Structural Chemistry, Fujian Institute of Research on the Structure of Matter, Chinese Academy of Sciences, Fuzhou, Fujian 350002, People's Republic of China}
\affil[13]{Department of Electrical and Computer Engineering, University of California, Riverside, CA, 92521 USA}
\affil[14]{Collaborative Innovation Center of Advanced Microstructures, 210093, Nanjing, Jiangsu, China}

\affil[$*$]{These authors contributed equally}
\affil[$\dagger$]{Corresponding authors: 
\textcolor{blue}{jiongy@t.shu.edu.cn, jma3@sjtu.edu.cn, huixing@sjtu.edu.cn, fmg12@cam.ac.uk, b.matteo@sjtu.edu.cn}}

\begin{abstract}
\textbf{Nowotny chimney ladder (NCL) crystals present physical properties in between the contrasting paradigms of ideal crystal and amorphous solid, making them promising candidates for thermoelectric applications due to their inherently low thermal conductivity. In this work, we report an extensive experimental characterization of the thermodynamic and thermoelectric transport properties of a large class of NCL materials, focusing on the intermetallic compound Ru$_2$Sn$_{3}$. We show that, despite their ordered crystalline structure, the heat capacity of these NCL compounds deviates from the Debye model at low temperatures and exhibits a boson-peak-like glassy anomaly in the range of $\mathbf{8}$-$\mathbf{14}$ K. By combining experimental measurements with density functional theory (DFT) and \emph{ab initio} molecular dynamics (AIMD) simulations, we attribute the microscopic origin of this glassy behavior to extremely low-energy optical phonons that universally emerge from the chimney ladder sublattice structure. Crucially, their coupling to acoustic phonons induces hybridization and avoided crossings, leading to strongly modified acoustic modes that directly contribute to the anomaly as well, similar to the case of other thermoelectric materials such as clathrates. Additionally, the measured thermal conductivity and the thermoelectric response present distinct anomalous glass-like features that strongly correlate with the dynamics of the low-lying optical phonons revealed by simulations. In particular, the electric resistivity displays an extended linear in $T$ behavior and an anomalously large $T^2$ contribution at low temperature. We propose a simple theoretical framework, based on electrons scattering with overdamped phononic modes, that qualitative explains both these features. Our work demonstrates that low-energy optical modes in ordered crystals can induce glassy behavior, outlining a pathway to design metallic materials with low thermal conductivity and unique thermoelectric properties without the need for disorder or strong electronic correlations.}
\end{abstract}

\makeatletter
\DeclareRobustCommand{\hbar}{{\mathchar'26\mkern-9muh}}
\makeatother
\DeclareUnicodeCharacter{2212}{\ensuremath{-}}

\begin{document}
\flushbottom
\maketitle
\thispagestyle{empty}

\section*{Introduction}
The theoretical description of vibrational, thermodynamic, and transport properties in solids has traditionally been structured around two extremes: the idealized ordered crystal and its opposite, the fully disordered amorphous solid. While the former is successfully rationalized by well-established paradigms such as Debye theory and Fermi liquid theory \cite{kittel2018introduction}, the latter is known to present characteristic glassy anomalies that are usually ascribed to the underlying structural disorder \cite{doi:10.1142/q0371}. 

Complex crystals present an intricate interplay of order and disorder at different length scales that might lead to the coexistence of crystal-like and glass-like features. In fact, since the first observation in 1981 \cite{PhysRevB.23.3886}, there are now many known examples of crystals with minimal or even absent structural disorder that display glass-like features, such as a boson-peak-like anomaly in the heat capacity normalized by Debye law, $C(T)/T^3$, or extremely low-thermal conductivity with a robust plateau in the same temperature range. Examples of crystalline systems exhibiting glassy anomalies can be found in ferroelectric materials \cite{doi:10.1080/00150199508007846,PhysRevMaterials.3.084414,2021arXiv210401969I}, superionic conductors \cite{KOBAYASHI1996460}, incommensurate crystals \cite{PhysRevLett.76.2334,PhysRevLett.114.195502,PhysRevLett.93.245902,PhysRevB.70.212301}, thermoelectrics \cite{PhysRevLett.100.165503,PhysRevB.74.125109,RevModPhys.86.669,doi:10.1073/pnas.1410349111,AKHBARIFAR2020128153,C3TA14021K,D1EE00738F,10.1063/1.2831926,C9TC04023D,B202464K,Christensen2008,Delaire2011,Lanigan-Atkins2020,Li2015,Tse2005,Ma2013}, clathrates~\cite{PhysRevLett.104.018301,Stefanoski_2010,PhysRevLett.82.779,Sigma}, and more \cite{Ren2021,PhysRevB.99.024301,Wu2024,PhysRevB.74.060201,PhysRevLett.78.82,PhysRevB.58.745,PhysRevB.66.012201,10.1063/1.4929530,PhysRevLett.118.105701,PhysRevB.85.134201,PhysRevB.97.201201,PhysRevB.110.174204,PhysRevLett.133.216101}.

The appearance of these anomalies in crystalline systems indicates that structural disorder is a sufficient but not necessary ingredient for the emergence of glassy physics and calls for alternative explanations that do not rely on it. The most promising proposal to explain the microscopic origin of glassiness in complex crystals invokes the presence of exceptionally low-energy and numerous optical modes \cite{PhysRevLett.93.245902,PhysRevB.70.212301,PhysRevB.99.024301,krivchikov2022role,doi:10.1142/S0217979221300024,Schliesser_2015,Baggioli_2020}. Following this original idea proposed by Ackermann \cite{PhysRevB.23.3886}, Cano et al. extended the same concept to incommensurate phases \cite{PhysRevLett.93.245902,PhysRevB.70.212301}. The first clear demonstration that low-energy optical modes can induce glassy features was presented in \cite{PhysRevB.99.024301}, using a molecular crystal, CCl$_4$. Around the same time, similar conclusions were drawn in \cite{PhysRevMaterials.3.084414,2021arXiv210401969I}, where the focus was on ferroelectric structural instabilities. Aside from the microscopic details, which are certainly different, this scenario shows strong similarities to the recently reported boson-peak flat band, discussed and experimentally observed in amorphous solids \cite{Hu2022,PhysRevResearch.5.023055,Tomterud2023,PhysRevLett.133.188302,Jiang_2024,mahajan2026flatbandperspectivebosonpeak,mizuno2025bosonpeakcovalentnetwork}.

Despite some preliminary positive tests of the low-energy optical phonon hypothesis in minimally disordered systems, a convincing proof is still missing and its universality has still to be verified. Moreover, since only a few of the so far explored compounds are metallic, the impact of these glass-like modes on thermoelectric transport properties and the interplay between glassiness and electronic transport in complex crystals have not been investigated yet, rendering impossible a comparison to the anomalous thermoelectric properties reported in metallic glasses \cite{HOWSON1988265,pryadun2015thermoelectric,PhysRevB.74.014208}.

In this work, we examine a large set of materials from the Nowotny chimney ladder (NCL) family \cite{schwomma63,lu02}, intermetallic crystals with weakly coupled helical sublattices, with particular focus on the case of Ru$_2$Sn$_{3}$. The intricate NCL structure can generate exotic ground states ranging from chiral ferromagnetism \cite{han_anisotropic_2016} to one-dimensional topological surface states \cite{gibson_quasi_2014}, and a very rich and complex vibrational spectrum. Moreover, extremely low and glassy thermal conductivity has been reported in some of these materials \cite{Kawasoko_2014,Chen2015} and associated with the existence of low-lying optical modes arising from the sliding of the helical sublattices and the large $c$-axis \cite{Chen2015}.

In summary, NCL materials are ideal candidates for exploring emergent glassy behavior within simple crystal structures, with the added potential to extend this analysis to thermoelectric transport properties due to their metallic nature. This work aims to determine whether these crystals display glassy thermodynamic and transport properties, and, if so, to uncover the microscopic origin of this behavior.

\begin{figure}[h]
    \centering
    \includegraphics[width=\linewidth]{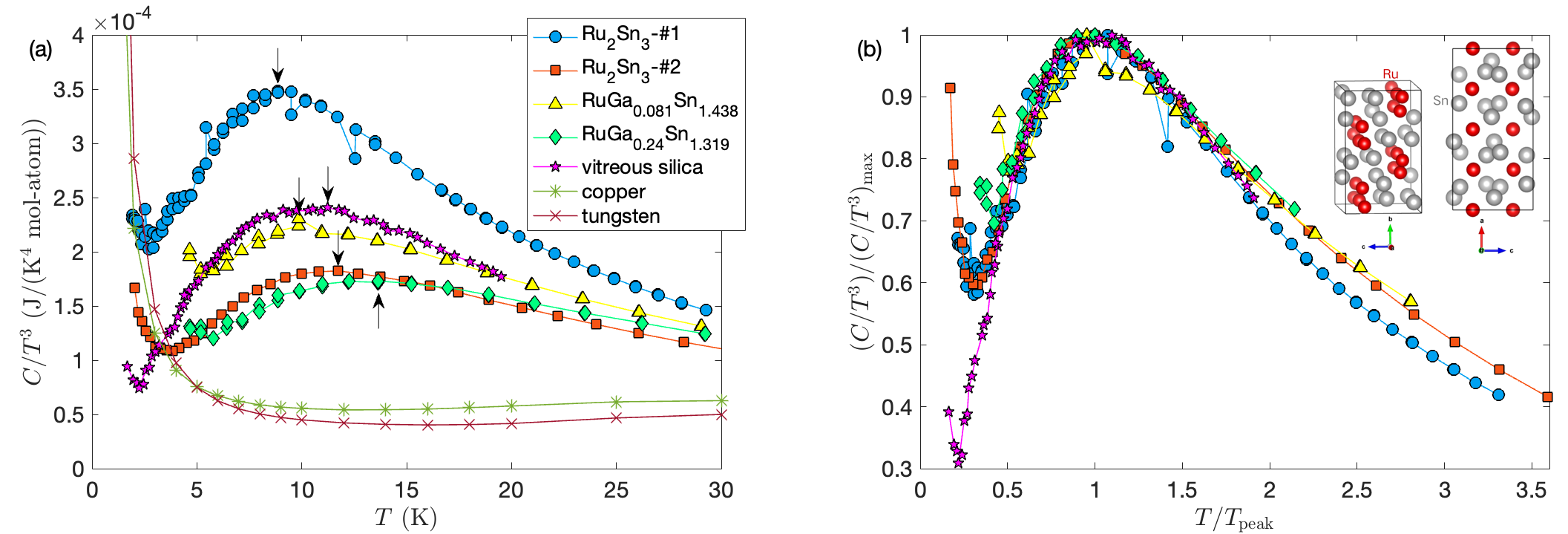}
    \caption{\textbf{(a)} Measured heat capacity $C(T)$ normalized by $T^3$ for several NCL materials: Ru$_2$Sn$_{3}\text{-}\#1$, Ru$_2$Sn$_{3}\text{-}\#2$, RuGa$_{0.081}$Sn$_{1.438}$, RuGa$_{0.24}$Sn$_{1.319}$. The experimental data for vitreous silica, copper, and tungsten (taken with permission from Refs.~\cite{PhysRevB.34.5665,10.1063/1.555728}) are also reported for direct comparison. The black arrows indicate the location of the boson-peak-like anomaly. $\#1$ and $\#2$ refer to the two different samples synthesized (see SM for details). \textbf{(b)} Same experimental data normalized by the position and the intensity of the boson-peak-like anomaly. The universal collapse of the experimental curves, aside from the very low temperature limit, is evident. The inset provides a schematic view of the orthorhombic chimney–ladder crystal structure of Ru$_2$Sn$_{3}$, shown along different crystallographic axes.}
    \label{fig:1}
\end{figure} 

\section*{Glassy heat capacity and its physical origin} 
In Fig.~\ref{fig:1}(a), we report the experimental results for the heat capacity $C(T)$ of Ru$_2$Sn$_{3}$ and other NCL materials divided by $T^3$. More details on the synthesis and characterization of the materials can be found in the Supplementary Material (SM). We stress that x-ray diffraction analysis and Rietveld refinements (see SM) confirm that our Ru$_2$Sn$_3$ samples are fully crystalline and that disorder effects are negligible. The characteristic chimney--ladder crystal structure of Ru$_2$Sn$_3$ is shown in the inset of Fig.~\ref{fig:1}(b), viewed along different crystallographic axes, highlighting the sublattice arrangement along the $c$-axis.

We observe that all the NCL materials exhibit an upturn of $C/T^3$ at low temperatures, which is due to the Sommerfeld electronic contribution $C_{\text{el}}= \gamma T$ \cite{kittel2018introduction}. We notice that the upturn in the experimental data for vitreous silica (SiO$_2$) is not of the same nature, since this glass is a good insulator, but rather a manifestation of two-level states (TLS) typical of amorphous solids \cite{doi:10.1142/q0371}. More importantly, all the compounds reveal a clear excess peak over the Debye law ($C_{\text{Debye}}\propto T^3$), strongly resembling the boson peak anomalies seen in amorphous compounds \cite{doi:10.1142/q0371}. The location of this peak varies with the chemical composition of the compound in a range between $8$ K and $14$ K, and it is compatible with that of SiO$_2$, a prototypical amorphous solid (displayed for comparison in Fig.~\ref{fig:1}(a)). 

More precisely, we find the excess peak at the following positions in each compound:
\begin{center}
\begin{tabular}{l l}
Ru$_2$Sn$_{3}$-\#1: & $T_{\text{peak}} = 8.8\,\mathrm{K}$; \\
Ru$_2$Sn$_{3}$-\#2: & $T_{\text{peak}} = 11.7\,\mathrm{K}$; \\
RuGa$_{0.081}$Sn$_{1.438}$: & $T_{\text{peak}} = 9.9\,\mathrm{K}$; \\
RuGa$_{0.24}$Sn$_{1.319}$: & $T_{\text{peak}} = 13.6\,\mathrm{K}$; \\
vitreous silica \cite{PhysRevB.34.5665}: & $T_{\text{peak}} = 11.2\,\mathrm{K}$.
\end{tabular}
\end{center}
In contrast, ideal crystalline metals, such as copper and tungsten (displayed for comparison in Fig.~\ref{fig:1}(a)), show perfect agreement with Debye's law in the same temperature range. 

\begin{figure}[ht]
    \centering
    \includegraphics[width=\linewidth]{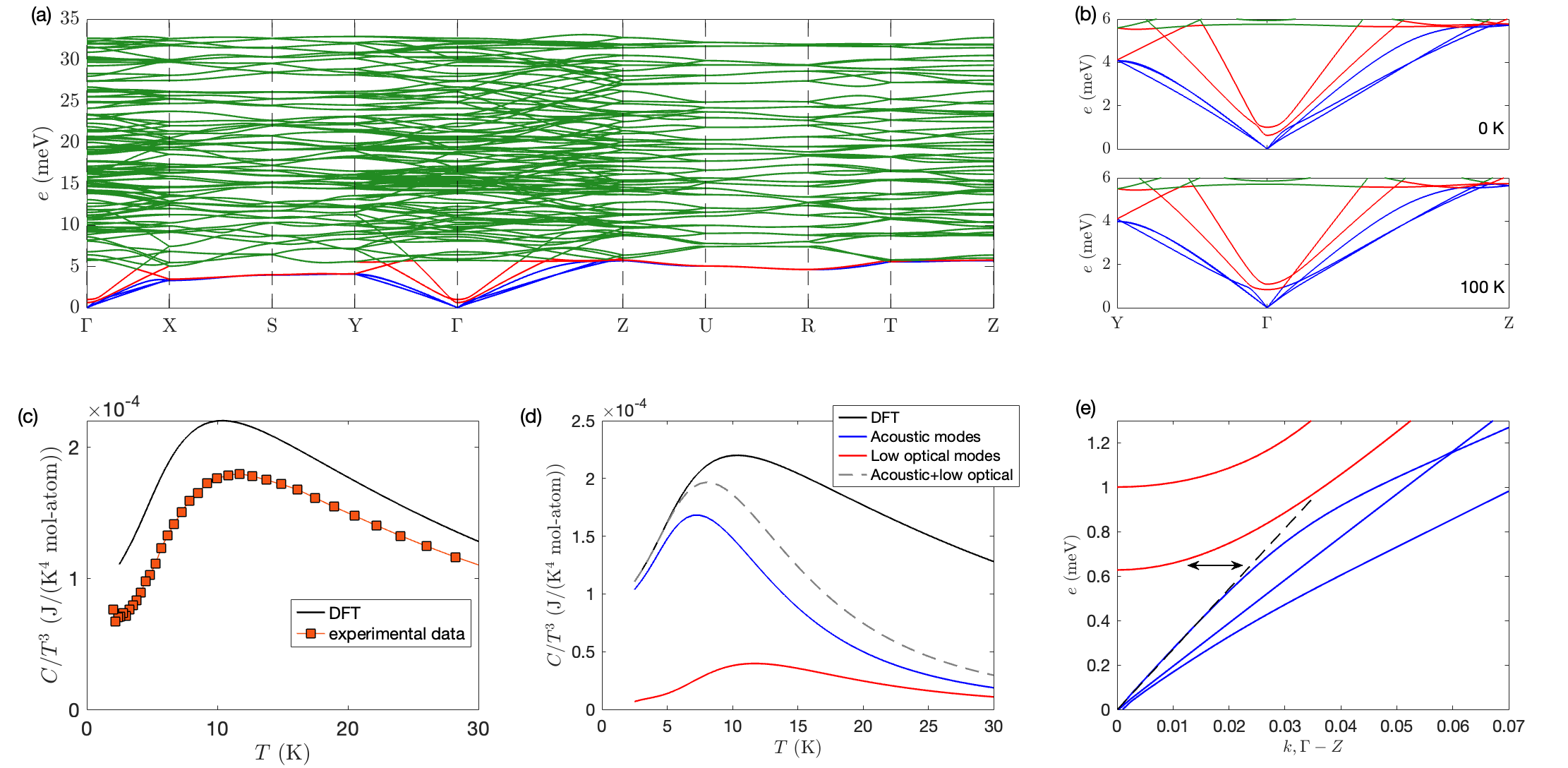}
    \caption{\textbf{(a)} Vibrational spectrum of Ru$_2$Sn$_3$ across the whole Brillouin zone obtained from zero-temperature DFT simulations. \textbf{(b)} A comparison of the low-energy spectrum around the $Y-\Gamma-Z$ directions from zero-temperature ($0$ K) DFT simulations (top panel) and \emph{ab initio} molecular dynamics (AIMD) results at $100\,\mathrm{K}$ (bottom panel). See SM for further discussion of the anharmonic effects and a more detailed comparison of the whole vibrational spectrum. \textbf{(c)} Comparison between the theoretical estimate of the heat capacity from DFT and the experimental data (with electronic contribution subtracted) for sample Ru$_2$Sn$_3$-\#2. 
\textbf{(d)} Mode resolved theoretical computation of the heat capacity from DFT. The mode-resolved contributions are shown using the same color scheme as in panels (a)-(b), and they are limited to the five lowest modes.  \textbf{(e)} Zoom of the low-energy region, highlighting acoustic modes (blue curves) and the two branches of anomalously low-lying optical modes along the $\Gamma$–$Z$ direction (red curves). The interaction and avoided crossing between the modes, in particular the lowest optical mode and the highest acoustic one (horizontal double arrow), are evident. The black dashed line denotes the ideal linear dispersion $e\propto k$ of the acoustic mode with the largest group velocity, highlighting its deviation from this ideal behavior at energies around $0.5$ meV, close to that of the lowest optical mode at the $\Gamma$ point. } 
    \label{fig:2}
\end{figure}

In Fig.~\ref{fig:1}(b), we show the same experimental data after normalizing them to the intensity and the position of the peak ($T_{\text{peak}}$). Guided by the original ideas of Refs.~\cite{PhysRevB.99.024301,krivchikov2022role,Miyazaki}, we find that all experimental curves collapse onto a single curve over the interval $0.5\,T_{\text{peak}} < T < 1.8\,T_{\text{peak}}$. The discrepancy below $0.5\,T_{\text{peak}}$ arises from electronic contributions (or, in the case of vitreous silica, from two-level systems) that are not subtracted from the experimental data. In contrast, deviations above $\approx 2\,T_{\text{peak}}$ originate from high-energy optical modes, which reflect the microscopic properties of each material and therefore vary across compounds. This feature is illustrated in Fig.~\ref{fig:2}(a) for Ru$_2$Sn$_3$-\#1, where a band of numerous, nearly flat optical modes (green colored curves) appears at energies approximately twice those of the two low-lying modes (red curves in Fig.~\ref{fig:2}(a)). In summary, the collapse around the heat-capacity peak, evident in Fig.~\ref{fig:1}(b), supports the ideas of Refs.~\cite{PhysRevB.99.024301,krivchikov2022role,Miyazaki} and points toward a universal description of this phenomenon that transcends structural differences, both among these materials and between crystals and glasses.

For completeness (see details in the SM), we have performed a low-temperature analysis of the data using the common fitting function $C(T)=a T+ b T^3$. For the two Ru$_2$Sn$_3$ samples, we find respectively $a=[2.35,3.73]$ $10^{-4}$ J/K$^2$ mol-atom and $b=[1.72,0.716]$ $10^{-4}$ J/K$^4$ mol-atom. From there, we can evaluate the Debye temperature of each compound using $\Theta_D=(\frac{12\pi^4}{5}\frac{N_{mol}k_B}{b})^{1/3}$. This allows us to verify whether the shift in $T_{\text{peak}}$ is just a consequence of the difference density, and therefore Debye temperature, of each compound. In the SM, we show the same experimental data after normalizing the $x$-axes by the Debye temperature. Doing so, the position of the peak for the two Ru$_2$Sn$_3$ samples is nearly the same, as expected. On the other hand, the density change encoded in $\theta_D$ is not enough to account for the energy shift of the peak in the Ga-doped samples. This is consistent with the results presented in Ref. \cite{Miyazaki}.

From a theoretical perspective, we notice that, due to the crystalline nature of our samples, the glass-like excess in the heat capacity cannot be attributed to the presence of structural disorder as assumed in the typical theoretical frameworks for glasses \cite{doi:10.1142/q0371} (\textit{e.g.}, soft potential model or heterogeneous elasticity theory). Therefore, to gain better insight on the microscopic origin of the boson-peak-like anomaly in NCL materials, we resort to density functional theory (DFT) and \emph{ab initio} molecular dynamics (AIMD) simulations, and we focus on the Ru$_2$Sn$_{3}$ compound.

Fig.~\ref{fig:2}(a) shows the vibrational spectrum of Ru$_2$Sn$_3$, obtained from zero-temperature DFT simulations across the entire Brillouin zone up to energies of $\approx 35\,\mathrm{meV}$.

In blue, we observe the dispersion of the acoustic phonon modes. In red, we highlight two exceptionally low-energy optical branches with energies comparable to those of the acoustic modes. These modes were further analyzed using animation software to identify their vibrational patterns (see SM for details).

The first low-energy optical branch corresponds to a corkscrew-like motion of the Sn helices through the Ru lattice, accompanied by a rigid counter-motion of the Ru sublattice. This mode is consistent with the previously reported “twisting” mode in Mn$_4$Si$_7$ \cite{Chen2015}. The first low-energy optical branch corresponds to a corkscrew-like motion of the Sn helices inside the Ru framework, accompanied by a rigid counter-motion of the Ru sublattice relative to the Sn sublattice. This mode is consistent with the low-frequency twisting motions of the Si ladders inside the Mn chimneys reported in Mn$_4$Si$_7$ \cite{Chen2015}. Similar dispersions have also been observed in GaN bulk crystals due to lattice anisotropy \cite{Zhang2024}. 

The second branch corresponds to a canting motion of Sn atoms within a relatively rigid Ru host. This mode involves oscillations between the atomic positions of the low-temperature orthorhombic phase and those of the high-temperature tetragonal phase \cite{poutcharovsky_diffusionless_1975}. It is associated with a structural phase transition occurring at higher temperatures ($\approx 150$–$200$ K, depending on composition). Notably, this mode is highly dispersive and exhibits a group velocity larger than that of the acoustic modes (see Fig.~\ref{fig:2}(b)). Additional details on these low-lying modes are provided in the SM.

Beyond these branches, the vibrational spectrum exhibits a large number of quasi-localized, nearly dispersionless optical modes (shown in green), with energies between $6$ and $35\,\mathrm{meV}$. These modes originate from the structural complexity of Ru$_2$Sn$_3$ and its large unit cell.

To assess the stability of the spectrum and the role of anharmonic effects, we performed AIMD simulations at finite temperature. A complete comparison of the vibrational spectrum at $0$ K and $100$ K across the whole Brillouin zone is provided in the SM. A zoom along the $Y-\Gamma-Z$ directions is shown in Fig.~\ref{fig:2}(b). The top panel shows the zero-temperature dispersions obtained from DFT, while the bottom panel presents the corresponding results at $100$ K from AIMD.

Overall, the phonon dispersions are only weakly affected by finite temperature. As a result, the modifications observed at $100$ K are not expected to significantly impact the heat capacity below $30$ K, consistent with Fig.~\ref{fig:2}(c). In contrast, these effects become important for transport properties, particularly for the thermal conductivity at higher temperatures.

To further quantify anharmonic effects, the Supplementary Material presents additional calculations of the phonon group velocity and scattering rate as functions of frequency. These results will be discussed in detail in connection with the thermal conductivity measurements.

To continue, we compute the mode-resolved heat capacity within the harmonic approximation,
\begin{equation}
    C_v(T)=\frac{N_A k_B}{N_{\mathbf{q}}} \sum_{\mathbf{q}}\left(\frac{\hbar \omega_{\mathbf{q} \nu}}{k_B T}\right)^2 \frac{\exp \left(\frac{\hbar \omega_{\mathbf{q} \nu}}{k_B T}\right)}{\left[\exp \left(\frac{\hbar \omega_{\mathbf{q} \nu}}{k_B T}\right)-1\right]^2}, \label{eqs}
\end{equation}
where $N_A$ is the Avogadro number, $k_B$ the Boltzmann constant, $N_{\mathbf{q}}$ the number of sampled $\textbf{q}$-points and $\omega_{\mathbf{q} \nu}$ the phonon frequency per mode. The total heat capacity is then directly obtained by summing over all modes,
\begin{equation}
    C(T)=\sum_v C_v(T).
\end{equation}
In the following, we compute the heat capacity using the formulas above and the vibrational spectrum obtained from DFT at zero temperature (see Fig.~\ref{fig:2}(a)).

A comparison between the theoretical estimate and the experimental data is shown in Fig.~\ref{fig:2}(c). Overall, the calculation reproduces the temperature dependence of the heat capacity remarkably well, in particular capturing the boson-peak-like excess around $11$ K. However, it slightly overestimates the absolute magnitude of $C(T)$ over the entire temperature range.

In order to better understand the microscopic origin of the boson-peak-like excess observed in the experimental data around $11$ K, we computed the mode-resolved heat capacity from the DFT spectrum, focusing on the contributions of the acoustic modes (blue) and the two lowest optical modes (red), as shown in Fig.~\ref{fig:2}(d). We first observe that the low-lying optical modes—highlighted in red in Fig.~\ref{fig:2}(a)-(b)—produce a pronounced peak whose temperature closely matches the anomalous feature in the total heat capacity. This contribution (red curve in Fig.~\ref{fig:2}(d)) mainly originates from their small energy gap and nearly flat dispersion near the $\Gamma$ point.

At the same time, the acoustic modes (blue curve) also display an excess over the Debye level, with a peak at slightly lower temperature. This indicates that the total anomaly around $11$ K arises from a combined effect: the low-energy optical modes and the strongly distorted acoustic modes both play a role.

Finally, we observe that the combined contribution of these five modes accounts for a large portion of the excess peak in the heat capacity. However, it does not fully reproduce its shape, particularly the high-temperature tail. In that regime, higher-energy optical modes become increasingly relevant and must be included to recover the full behavior, as shown in Fig.~\ref{fig:2}(c).

To understand why acoustic modes can produce an excess over the Debye contribution, we examine their dispersion near the $\Gamma$ point. As a concrete example, Fig.~\ref{fig:2}(e) shows a zoom along the $\Gamma$–$Z$ direction; similar features are observed along different directions as shown in panel (a) of the same Figure. Two of the acoustic branches—those with the lowest group velocity near $\Gamma$—remain nearly linear up to relatively high energies. As such, they are expected to contribute primarily a standard Debye-like behavior, $C \sim T^3$, with minimal impact on the excess peak.

In contrast, the third acoustic branch, which has the largest group velocity near $\Gamma$, exhibits a pronounced anomaly. This behavior originates from its interaction with the lowest optical mode, which has an extremely small gap at $\Gamma$ ($\approx 0.6$ meV). The coupling between these two modes leads to a clear avoided crossing and a strong deviation of the acoustic dispersion from the expected linear trend (black dashed line in Fig.~\ref{fig:2}(e)). This avoided crossing, together with the associated softening of the acoustic mode near $ \approx 0.6$ meV, is responsible for the peak observed in the acoustic contribution in Fig.~\ref{fig:2}(d).

In summary, our analysis indicates that the boson-peak-like anomaly in the heat capacity originates primarily from the presence of extremely low-lying optical modes and their coupling to the acoustic branch with the largest group velocity. This interplay leads to a pronounced avoided crossing and a strongly distorted acoustic dispersion, which together with the lowest optical modes constitute the main source of the observed glass-like feature.

We emphasize that the presence of extremely low-lying optical modes and glass-like features have already been reported in other chimney ladder compounds, most notably Mn$_4$Si$_7$ \cite{Chen2015}. This points to the possibility that a boson-peak-like excess, originating from low-lying optical modes, may be a universal characteristic of chimney ladder structures.

In this context, it is worth highlighting the strong parallels with other crystalline systems that exhibit glass-like behavior, particularly thermoelectric materials such as clathrates \cite{PhysRevB.77.235119} and related compounds \cite{doi:10.7566/JPSJ.84.103601}. In these systems, rattling modes and avoided-crossing phenomena \cite{Christensen2008,10.1063/1.475218}—hallmarks of the so-called phonon-glass paradigm \cite{RevModPhys.86.669}—give rise to glass-like anomalies in both the heat capacity and the thermal conductivity (see \cite{PhysRevB.100.220201} for a theoretical framework for the boson-peak excess). From a microscopic perspective, the key distinction is that in those systems the avoided crossing and low-lying optical modes originate from the host–guest structure and the resulting rattler excitations. In contrast, in chimney ladder materials these features emerge from the peculiar sublattice arrangement and the associated soft sliding modes, more similar to the case of incommensurate phases \cite{PhysRevLett.93.245902,PhysRevB.70.212301}.

Finally, we observe a noticeable variation in $C(T)$ between the two Ru$_2$Sn$_3$ samples, which becomes less pronounced after proper Debye normalization (see SM). Returning to the doped samples in Fig.~\ref{fig:1}, partial substitution of Sn with Ga shifts the excess peak to higher temperatures while reducing its intensity. Despite the dramatic structural effect—where even $4\%$ Ga increases the $c$-axis by an order of magnitude (see SM)—this trend indicates that incommensuration is not the primary factor controlling the gap of the low-energy modes. This unusual lattice expansion deviates from Vegard’s law and reflects the weak coupling between the Ru and Sn sublattices. As a Nowotny chimney–ladder phase, the system consists of two interpenetrating sublattices with largely independent, often incommensurate periodicities; Ga substitution modifies their mismatch, leading to large changes in the axial repeat distance. Finally, as shown in the SM and consistent with Ref.~\cite{Miyazaki}, changes in density alone cannot account for the observed shift of the boson-peak energy.

\section*{Glass-like thermal conductivity}
\begin{figure}[ht]
    \centering
\includegraphics[width=\linewidth]{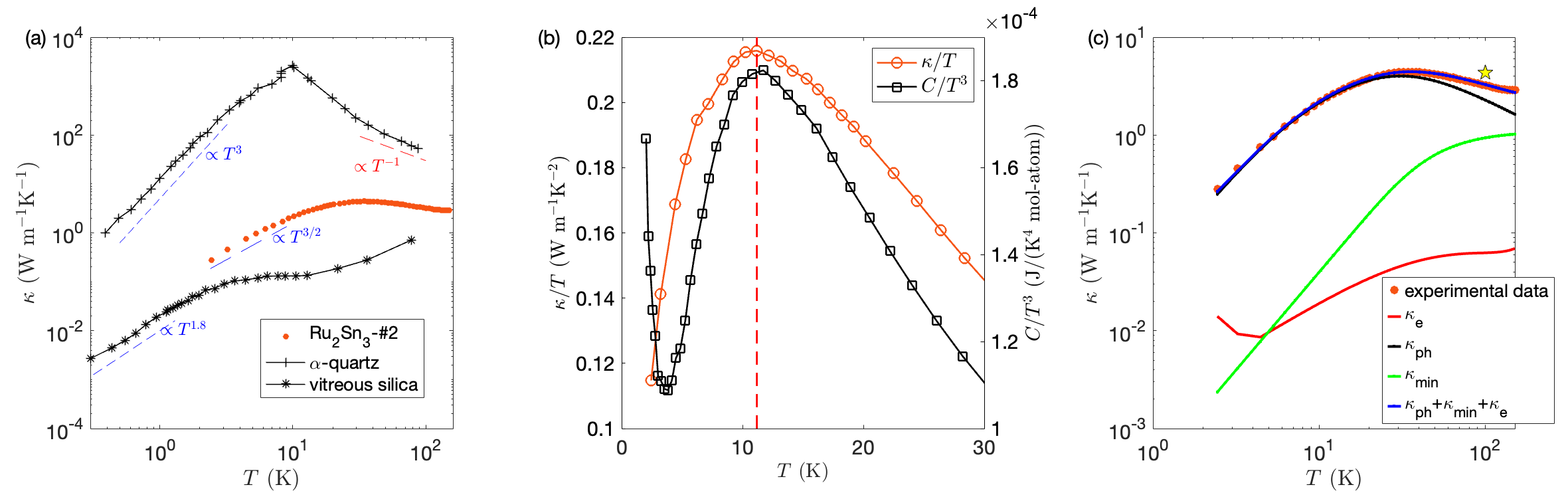}
\caption{\textbf{(a)} Measured thermal conductivity $\kappa(T)$ of Ru$_2$Sn$_{3}\text{-}\#2$. Data for $\alpha$-quartz and vitreous silica are reproduced from Ref.~\cite{PhysRevB.4.2029}. \textbf{(b)} Reduced thermal conductivity $\kappa(T)/T$ and reduced heat capacity $C(T)/T^3$ of Ru$_2$Sn$_{3}\text{-}\#2$. The dashed line marks the peak position at $\approx 11.2$ K. \textbf{(c)} Theoretical analysis of the data based on the Debye–Peierls relaxation-time model ($\kappa_{\text{ph}}$), the Cahill–Pohl minimum thermal conductivity ($\kappa_{\text{min}}$), and the electronic contribution ($\kappa_e$) computed using the Wiedemann–Franz law. The yellow star indicates the calculated lattice thermal conductivity at 100 K using AIMD (See more details in SM, Section~\ref{SM:AIMD}). A linear-linear representation of the measured thermal conductivity is provided for completeness in SM.} 
    \label{fig:3}
\end{figure}

Anomalies in thermal conductivity, relative to the textbook behavior of ideal crystals, are another hallmark of glassy physics \cite{PhysRevB.4.2029}. In Fig.~\ref{fig:3}(a), we present the measured thermal conductivity of Ru$_2$Sn$_3$. For comparison, we also show experimental data for an ideal crystal ($\alpha$-quartz) and an ideal glass (vitreous silica) from Ref.~\cite{PhysRevB.4.2029}, together with their characteristic low- and high-temperature limits.

We first note that the thermal conductivity $\kappa(T)$ of Ru$_2$Sn$_3$ is remarkably low—one to two orders of magnitude smaller than that of an ideal crystal—and comparable to typical glassy values. In addition, the phonon peak is weak and broad, in clear contrast to the sharp, well-defined peak observed in crystalline systems (see $\alpha$-quartz). Moreover, unlike minimally disordered crystals \cite{PhysRevLett.119.215506}, no clear plateau is observed.

Most strikingly, at low temperatures we find a scaling $\kappa \sim T^{3/2}$, which deviates strongly from the Debye prediction $\kappa \sim T^3$ for ideal crystals. Instead, it approaches the glassy behavior $\kappa \sim T^{2-\Delta}$ observed in materials such as vitreous silica (with $\Delta \approx 0.2$) \cite{doi:10.1142/q0371}. This anomalous regime coincides with the temperature range where the contribution of the two lowest optical modes to the heat capacity is maximal (see Fig.~\ref{fig:2}(d)), suggesting a direct connection between these modes and the thermal transport. Finally, at high temperatures we do not observe the characteristic $1/T$ decay associated with Umklapp scattering in crystals; instead, $\kappa(T)$ exhibits a weaker decrease, likely related to the presence of numerous nearly dispersionless optical modes (green curves in Fig.~\ref{fig:2}(a)).

Before introducing a theoretical model for the thermal conductivity, we note that $\kappa$ provides a complementary self-consistency check of the boson-peak-like anomaly discussed above. In general, the total thermal conductivity is the sum of phononic (p) and electronic (e) contributions, $\kappa = \kappa_e + \kappa_p$. Since both terms scale with the specific heat of their respective carriers, $C_{e,p}$, and are weighted by their velocities $v_{e,p}$ and mean free paths $l_{e,p}$, one can write $\kappa \propto C_e v_e l_e + C_p v_p l_p$. In the simple limit where $v$ and $l$ are only weakly temperature dependent (as is often the case at low temperatures), the thermal conductivity effectively tracks the temperature dependence of the specific heat.

In Fig.~\ref{fig:3}(b), we therefore analyze $\kappa/T$, where the $1/T$ factor removes the linear temperature dependence of the electronic specific heat ($C_e \propto T$), allowing the phononic contribution to be more clearly resolved. Strikingly, a pronounced broad peak in $\kappa/T$ appears at $T \approx 11.2$ K, closely matching the peak in $C/T^3$ ($\approx 11.7$ K) shown in Fig.~\ref{fig:1}. Since thermal conductivity selectively probes heat-carrying excitations, unlike specific heat which includes all degrees of freedom, the coincidence of these peak temperatures provides strong evidence for a boson-peak anomaly. By contrast, a $T^{3}$ normalization is not suitable for $\kappa$ in metallic Ru$_2$Sn$_3$, since it would disproportionately enhance the low-temperature electronic contribution, thereby rendering the broad glass-like anomaly much less discernible.

 \begin{figure}
    \centering
\includegraphics[width=\linewidth]{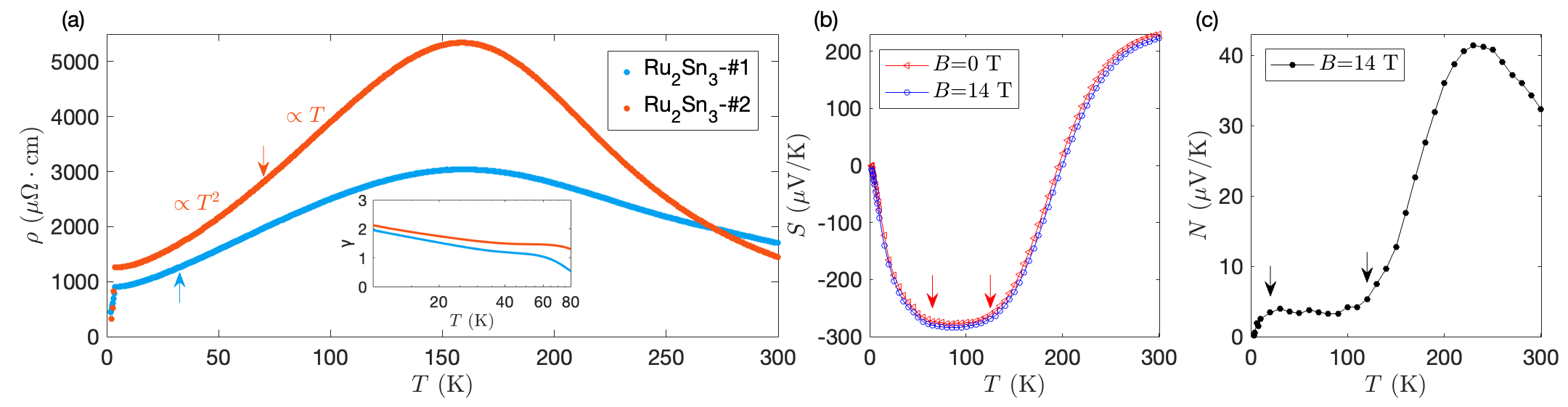}
\caption{\textbf{(a)} Measured resistivity $\rho(T)$ for Ru$_2$Sn$_{3}\text{-}\#1$ and Ru$_2$Sn$_{3}\text{-}\#2$. The arrows indicate the crossover temperatures between $T^2$ and $T$-linear region ($33$ K for Ru$_2$Sn$_{3}\text{-}\#1$ and $70$ K for Ru$_2$Sn$_{3}\text{-}\#2$). The inset shows the power-law exponent of the electric resistivity as a function of temperature. \textbf{(b)} Measured Seebeck $S(T)$ at $0$ T and $13.9$ T magnetic fields. The red arrows indicate the plateau region $60-120$ K. \textbf{(c)} Measured Nernst signal $N(T)$ at $13.9$ T magnetic field. The black arrows indicate the plateau region $20-120$ K.}
    \label{fig:4}
\end{figure} 

Following \cite{KRIVCHIKOV2015141}, we model the thermal conductivity using a combination of the Debye--Peierls relaxation-time approach \cite{peierls1955quantum} and the Cahill--Pohl model \cite{cpmodel}. Specifically, we write the total thermal conductivity as $\kappa = \kappa_{\text{ph}} + \kappa_{\text{min}} + \kappa_e$, where $\kappa_e$ denotes the electronic contribution. The Debye--Peierls term is given by
\begin{equation}
\kappa_{\text{ph}}(T)=\frac{k_B^4 T^3}{2 \pi^2 \hbar^3 c_s} \int_0^{\theta_D / T} \tau_R(x) \,\frac{x^4 e^x}{\left(e^x-1\right)^2} d x,
\end{equation}
where $\theta_D$ is the Debye temperature, $c_s$ the sound velocity, and $\tau_R(x)$ an effective relaxation time. The Cahill--Pohl contribution reads
\begin{equation}
\kappa_{\min }(T)=\frac{k_B^3 T^2}{2 \pi \hbar^2 c_s} \int_0^{\theta_D / T} \frac{x^3 e^x}{\left(e^x-1\right)^2} d x,
\end{equation}
with $x=\hbar \omega/(k_B T)$ in both expressions.

To fit the experimental data, we consider a total scattering rate of the form
\begin{equation}
\tau_R^{-1}(\omega,T)=\tau_U^{-1}+\tau_{\text{dis}}^{-1}+\tau_{\text{imp}}^{-1},
\end{equation}
where $\tau_U^{-1}=B \omega^2 T \exp(-E_u/T)$ describes Umklapp processes, $\tau_{\text{dis}}^{-1}=D_d \omega$ accounts for scattering by dislocations, and $\tau_{\text{imp}}^{-1}=C_{\text{imp}} \omega^4$ represents impurity scattering. 
Finally, the electronic contribution is evaluated using the Wiedemann--Franz law,
\begin{equation}
\kappa_e(T) = L_0 \, \sigma(T)\, T,
\end{equation}
where $L_0 = \frac{\pi^2}{3}\left(\frac{k_B}{e}\right)^2$ is the Lorenz number and $\sigma(T)$ denotes the electrical conductivity (see more details in SM).

In Fig.~\ref{fig:3}(c), we compare the theoretical model with the experimental data, finding excellent agreement. The best-fit parameters are $D_d=2.9\times10^{-3}$, $B=8.7\times10^{-17}$ s/K, $C_{\text{imp}}=4.35\times10^{-41}$ s$^3$, $\theta_D=200\,\mathrm{K}$, $E_u=\theta_D/3$, and $c_s=2060$ m/s. 

These results show that the Debye--Peierls contribution alone fails to capture the experimental trend, particularly at high temperatures. In contrast, in line with observations in molecular crystals \cite{KRIVCHIKOV2015141}, the combined model provides a substantially improved description, reproducing the data with good accuracy.

In Fig.~\ref{fig:3}(c), we also present a theoretical estimate of the thermal conductivity $\kappa$ at $T=100$ K obtained from AIMD simulations (see SM for details). Including anharmonic effects through three-phonon interactions, we obtain a lattice thermal conductivity of $4.31$ W m$^{-1}$ K$^{-1}$ (yellow star symbol in Fig.~\ref{fig:3}(c)), which is slightly higher than the experimental value of $3.22$ W m$^{-1}$ K$^{-1}$. This discrepancy is likely due to the neglect of additional scattering mechanisms, such as electron–phonon interactions and grain boundary scattering.

In the SM, we further analyze the frequency-dependent phonon group velocity and scattering rate. The maximum phonon group velocity reaches $5.27$ km s$^{-1}$; however, most phonon modes lie within the range $0$–$3$ km s$^{-1}$, particularly in the low- to medium-frequency regime that dominates heat transport. The scattering-rate-weighted average group velocity is only $0.60$ km s$^{-1}$, significantly lower than in high-$\kappa_l$ materials such as FeSi$_2$ ($1.26$ km s$^{-1}$ and $24.26$ W m$^{-1}$ K$^{-1}$ at 300 K) \cite{Zhao2017}.

Overall, the intrinsically low lattice thermal conductivity of orthorhombic Ru$_2$Sn$_3$ can be primarily attributed to its low average phonon group velocity.

\section*{Anomalous thermoelectric transport and a theoretical scenario}

The analysis of the thermal conductivity in Ru$_2$Sn$_3$ confirms that NCL crystals exhibit transport behavior intermediate between the ideal crystal and ideal glass limits. Taking advantage of the metallic nature of Ru$_2$Sn$_3$, we extend our study to thermoelectric transport by measuring the electrical resistivity $\rho(T)$, Seebeck coefficient $S(T)$, and Nernst signal $N(T)$.

In the Supplementary Material, we present the electronic band structure of Ru$_2$Sn$_3$ obtained from DFT at $T=0$ K, confirming the metallic nature of the compound. Figure~\ref{fig:4}(a) shows the measured resistivity for the two Ru$_2$Sn$_3$ samples. Both display a crossover from $T^2$ to $T$-linear behavior at characteristic temperatures of $\approx 33$ K ($2.6$ meV) for Ru$2$Sn${3}\text{-}\#1$ and $\approx 70$ K ($6$ meV) for Ru$2$Sn${3}\text{-}\#2$. Notably, the linear-in-$T$ regime is robust and extends to unusually low temperatures. Its energy scale correlates strongly with that of the flat optical modes, suggesting that electron scattering off these numerous low-energy modes underlies this behavior, as recently discussed in \cite{PhysRevB.109.235118} (see also \cite{PhysRevB.108.075168,Caprara2022} for related mechanisms involving overdamped phason modes).

Fitting the low-temperature resistivity to the Fermi-liquid form $\rho=\rho_0 + A T^2$ yields unusually large coefficients compared to typical metallic systems \cite{behnia2022}: $A=404$ n$\Omega$ cm/K$^2$ for Ru$_2$Sn$_3$\#2 and $A=402$ n$\Omega$ cm/K$^2$ for Ru$_2$Sn$_3$\#1. A comparable value, $A=360$ n$\Omega$ cm/K$^2$, is obtained for Mn$_4$Si$_7$ using data from Ref.~\cite{Chen2015} (see SM).

To further characterize this behavior, we analyze the resistivity using the phenomenological form $\rho(T)=\rho_0 + \alpha T^{\gamma}$. The extracted exponent $\gamma(T)$ is shown in the inset of Fig.~\ref{fig:4}(a). Remarkably, neither compound exhibits the conventional $T^5$ regime expected from standard electron–phonon scattering \cite{Mizutani_2001}; instead, the data are consistently described by a dominant and anomalously large $T^2$ contribution.

Additional insight is provided by Hall measurements, reported in the Supplementary Material. Within the plateau-like temperature window identified in the thermoelectric response (60–120 K; see Fig.~\ref{fig:4}(b)–(c)), both the Hall resistivity and Hall angle show only a weak temperature dependence and no anomalous features. This observation supports the view that the resistivity anomaly is not primarily driven by changes in the electronic band structure. Indeed, if band-structure effects were dominant, one would generally expect corresponding signatures in the Hall response.

Previous studies have shown that in amorphous and disordered crystalline alloys, a continuous dynamical structure factor can give rise to a $T^2$ low-temperature resistivity via electron–phonon interactions, replacing the standard $T^5$ behavior~\cite{Mizutani_2001,PhysRevLett.39.102,PhysRevB.16.2978}. Motivated by this analogy, we propose that a similar mechanism operates in our ordered system: overdamped phononic modes—arising here from low-lying optical excitations—can act as efficient scattering channels, leading to the observed $T^2$ contribution to the resistivity. A related scenario has recently been discussed in twisted bilayer graphene \cite{PhysRevB.108.075168}.

The theoretical description starts from the Green’s function of a single phononic mode with frequency $\omega$ and wave vector $q$,
\begin{equation}\label{gg}
    G(\omega,q)=\frac{1}{\omega^2-\Omega(q)^2+i \omega \Gamma(q)}.
\end{equation}
Here, $\Gamma(q)$ denotes the linewidth of the mode and $\Omega(q)$ its dispersion. For acoustic phonons \cite{chaikin1995principles}, one has $\Omega(q)=v q$ and $\Gamma(q)=D q^2$ at low $q$, where $v$ is the sound velocity and $D$ the damping coefficient. 
In general, $D$ may depend on temperature and on the underlying scattering mechanisms; here, we approximate it as temperature independent for simplicity. 
The corresponding spectral function is defined as $A(\omega,q)=-\frac{1}{\pi}\,\mathrm{Im}\,G(\omega,q)$ \cite{coleman2015}.

The electron--phonon contribution to the electrical resistivity can then be estimated using the Baym formula \cite{Mizutani_2001}. Assuming that the Fourier transform of the ionic potential, $U_p(q)$, is approximately constant, this expression can be written as
\begin{equation}\label{baym}
    \rho_{\text{e-ph}}(T)= \bar{\rho} \int_0^{q_{D}}\int_{0}^{\infty} q^3 \, a(\omega,q)\, \beta\, \omega \, n(\omega) \, d\omega \, dq,
\end{equation}
where $n(\omega)=\left(e^{\frac{\hbar \omega}{k_B T}}-1\right)^{-1}$ is the Bose--Einstein distribution, $a(\omega,q)=2\hbar q^2\big(n(\omega)+1\big)A(\omega,q)$ is the dynamical structure factor, and $q_D$ is the Debye wave vector. The prefactor $\bar{\rho}$ is an overall constant determined microscopically by quantities such as the atomic volume, ionic potential, Fermi velocity, and Fermi wave number (see \cite{Mizutani_2001}). 

For convenience, we introduce the Debye temperature $\Theta_D=\frac{\hbar v q_D}{k_B}$ and the damping temperature scale $T_D=\frac{\hbar D q_D^2}{k_B}$.
\begin{figure}[ht]
    \centering
\includegraphics[width=\linewidth]{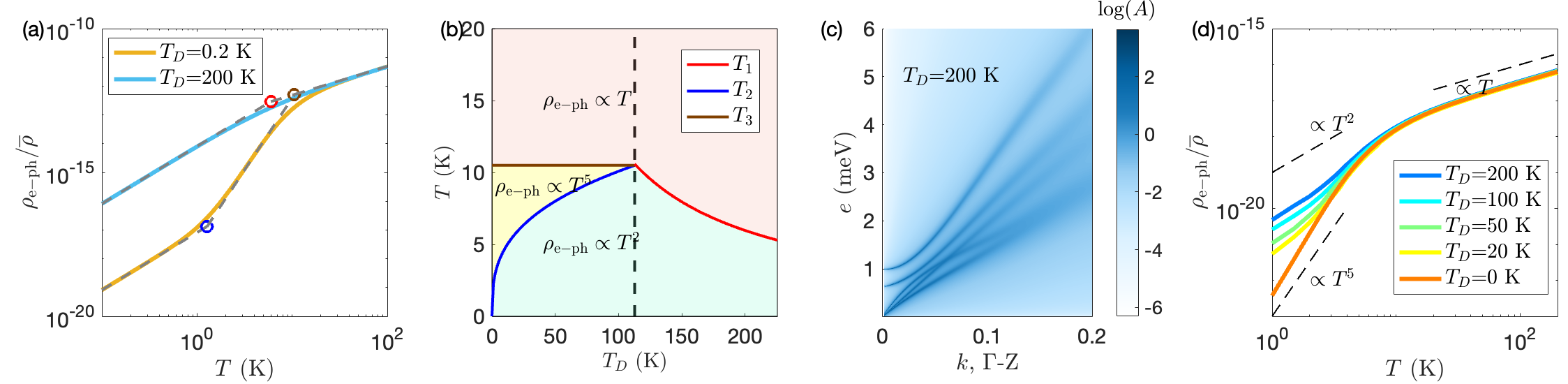}
\caption{\textbf{(a)} A test of the analytical expressions Eqs. \eqref{D-large}-\eqref{D-small} (dashed lines) with the numerical solutions of Eq. \eqref{baym} (solid lines). The two curves refer to the large and small damping regimes. The colored symbols indicate the temperature scales $T_1,T_2,T_3$ reported in the text. \textbf{(b)} The scaling behavior of $\rho_{\text{e-ph}}(T)$ in the $(T_D,T)$ plane. $\Theta_D$ is set to $50$ K in both panels. \textbf{(c)} Spectral function $A(k,\omega)$ using the spectrum in panel (c) of Fig. \ref{fig:2}. We set $T_D=200$ K. \textbf{(d)} Calculated electron-phonon contribution to the resistivity $\rho_{\text{e-ph}}(T)$ with different values of the damping parameter, $T_D$. }
    \label{SM:rhoT}
\end{figure}

In the large-damping regime, $T_D \gg \Theta_D$, we obtain
\begin{equation}\label{D-large}
\rho_{\text{e-ph}}(T) \approx \frac{\hbar \bar{\rho} q_D^6}{k_B}
\begin{cases}
\frac{T}{\Theta_D^{2}}\frac{1}{4}, & \text{if } T \gg T_{1}, \\
T^2 \frac{T_D}{\Theta_D^4} \frac{\pi}{6}, & \text{if } T \ll T_{1},
\end{cases}
\end{equation}
where $T_1=\frac{\Theta_D^2}{T_D}\frac{3}{2\pi}=\frac{3 \hbar}{2\pi k_B}\frac{v^2}{D}$ coincides with the crossover scale between underdamped and overdamped phonon dynamics.

In the small-damping regime, $T_D \ll \Theta_D$, we find
\begin{equation}\label{D-small}
\rho_{\text{e-ph}}(T) \approx \frac{\hbar \bar{\rho} q_D^6}{k_B}
\begin{cases}
T\frac{1}{\Theta_D^{2}}\frac{1}{4}, & \text{if } T \gg T_{3}, \\
124.4\, T^5 \frac{1}{\Theta_D^6}, & \text{if } T_2 \ll T \ll T_{3}, \\
T^2 \frac{T_D}{\Theta_D^4} \frac{\pi}{6}, & \text{if } T \ll T_2,
\end{cases}
\end{equation}
where $T_2 \approx (T_D \Theta_D^2)^{1/3}/6$ and $T_3 \approx \Theta_D/5$. When $T_D \approx 2.26\,\Theta_D$, all three characteristic scales coincide, $T_2=T_3=T_1$, marking the crossover between the two regimes. 

Figure~\ref{SM:rhoT}(a) confirms the validity of these analytical scalings through comparison with numerical evaluations of Eq.~\eqref{baym}, while Fig.~\ref{SM:rhoT}(b) summarizes the different regimes in the $(T_D,T)$ plane (note the similarities with \cite{PhysRevB.108.075168}).

Two key physical insights emerge from this analysis. (i) Scattering between electrons and overdamped phonons generates a $T^2$ contribution to the resistivity, which adds to the conventional electron--electron term. (ii) The high-temperature linear-in-$T$ regime can extend down to much lower temperatures, with the crossover scale controlled by the phonon damping and potentially much smaller than the Bloch--Grüneisen temperature.

We now apply this framework to Ru$_2$Sn$_3$. In Fig.~\ref{SM:rhoT}(c), we show the calculated spectral function including three acoustic modes and two low-energy optical modes, with $q=2\pi k/c$. The dispersions are taken from the DFT results in Fig.~\ref{fig:2}(c), and the damping scale is set to $T_D=200$ K. In Fig.~\ref{SM:rhoT}(d), we compute $\rho_{\text{e-ph}}(T)$ from Eq.~\eqref{baym} using this spectral function. The frequency integral is restricted to $\omega_{\text{max}} \approx 6$ meV to exclude the higher-energy optical modes (green curves in Fig.~\ref{fig:2}(a)). For simplicity, and to limit the number of free parameters, we assume a common damping for all branches, with $q_D=\frac{2\pi}{c}$ and $c=6.124\,\text{\AA}$. We note that, for optical modes, a $q$-independent damping term may also be relevant.

In the absence of damping ($D=0$), the acoustic contribution to $\rho_{\text{e-ph}}$ follows the standard Bloch--Grüneisen behavior, exhibiting a crossover from $T^5$ to linear-in-$T$ scaling \cite{Mizutani_2001}. For finite damping ($D \neq 0$), the low-temperature behavior instead becomes quadratic, $\rho_{\text{e-ph}} \sim T^2$, while the $T^5$ regime is suppressed.

Unfortunately, the phonon linewidths cannot be reliably extracted from our DFT calculations; we therefore treat the damping parameter $D$ as a free parameter in the model. Nonetheless, we find that physically reasonable values of $D$ yield a crossover between $T^2$ and linear-in-$T$ resistivity at temperatures comparable to those observed experimentally. This suggests that both the anomalous $T^2$ contribution and the extended linear-in-$T$ regime in Ru$_2$Sn$_{3}$ may originate from overdamped phonons. We speculate that this overdamped behavior of the acoustic modes arises from their coupling to low-lying optical modes, which enhances anharmonicity and scattering—an effect commonly observed in thermoelectric materials \cite{Jinfeng}.

A more quantitative description would require an accurate determination of the phonon linewidths, either experimentally or computationally (see SM for a preliminary analysis using AIMD). Inelastic neutron scattering provides the most direct experimental access, as it allows one to measure phonon damping under controlled conditions. However, such measurements require large, high-quality single crystals due to the limited neutron flux (as in, e.g., Mn$_4$Si$_7$ \cite{Chen2015}). Future work will therefore focus on optimizing the crystal growth of Ru$_2$Sn$_3$, either by producing larger single crystals or by co-aligning multiple samples, in order to enable a direct characterization of the phonon damping.

Finally, panels (b-c) of Fig.~\ref{fig:4} report experimental measurements of thermoelectric response in Ru$_2$Sn$_{3}\text{-}\#2$. Both the Seebeck coefficient and Nernst signal display an anomalous plateau in temperature.

The Seebeck coefficient exhibits an unusually large magnitude for a metallic system and displays a crossover from negative to positive values with increasing temperature. This behavior suggests that the electronic structure may deviate from that of a conventional metal with a large Fermi energy. Within a Fermi-liquid framework, the Seebeck coefficient is given by
\begin{equation}
S = -\frac{\pi^2}{3}\frac{k_{\mathrm{B}}^2 T}{e}
\left[
\frac{\partial \ln \tau(\epsilon)}{\partial \epsilon}
+
\frac{\partial \ln D(\epsilon)}{\partial \epsilon}
\right]_{\epsilon=\epsilon_{\mathrm{F}}},
\end{equation}
where $k_{\mathrm{B}}$ is the Boltzmann constant, $T$ the temperature, and $e$ the elementary charge. The quantity $\tau(\epsilon)$ denotes the energy-dependent carrier scattering time, while $D(\epsilon)$ is the electronic density of states. The derivatives are evaluated at the Fermi energy $\epsilon_{\mathrm{F}}$. The first term inside the brackets captures the contribution from the energy dependence of scattering processes, whereas the second term reflects the band-structure contribution and typically scales inversely with the Fermi energy. The large observed values of $S$ are therefore consistent with a small effective Fermi energy, as expected in a narrow-gap semiconductor or semimetallic scenario with strong conduction–valence band asymmetry.

At the same time, transport measurements, particularly the low-temperature resistivity, clearly indicate a metallic ground state. In this context, a semimetal with a small Fermi energy provides a plausible interpretation. However, the sign change of the Seebeck coefficient occurs in proximity to the structural transition near $160\,\mathrm{K}$, indicating that it cannot be solely attributed to a fixed band structure. In addition, the scattering contribution to $S$ is expected to be significant in this system due to strong interactions with low-lying optical phonons. These combined effects suggest that a quantitative description of the Seebeck response requires a more comprehensive theoretical treatment.

We note that the plateau is particularly robust in $N(T)$ and it strongly correlates with the energy scale of the lowest optical modes ($\approx 20$ K or $1.7$ meV) and the energy range in which numerous optical phonons appear ($\approx 60$ K or $5.2$ meV). This observation highlights the possibility of dominant electron-phonon scattering in the electron transport process. We advance the idea that the interaction between the electrons and the low-lying optical modes is the microscopic origin of the anomalous thermoelectric transport properties as well, although a more precise picture is beyond the scope of the present measurements. Along these lines, it is worth noting that recent studies have shown that electron-phonon scattering can strongly enhance the Seebeck coefficient~\cite{PhysRevX.15.021054,doi:10.1126/sciadv.adv7113}. Interestingly, a plateau in the Seebeck coefficient has been reported in Au-based and Cu-based metallic glasses \cite{pryadun2015thermoelectric, PhysRevB.74.014208} where it is associated with low-energy glassy excitations, suggesting another striking similarity between NCL compounds and structural glasses.

\begin{figure}[h]
    \centering
    \includegraphics[width=0.6\linewidth]{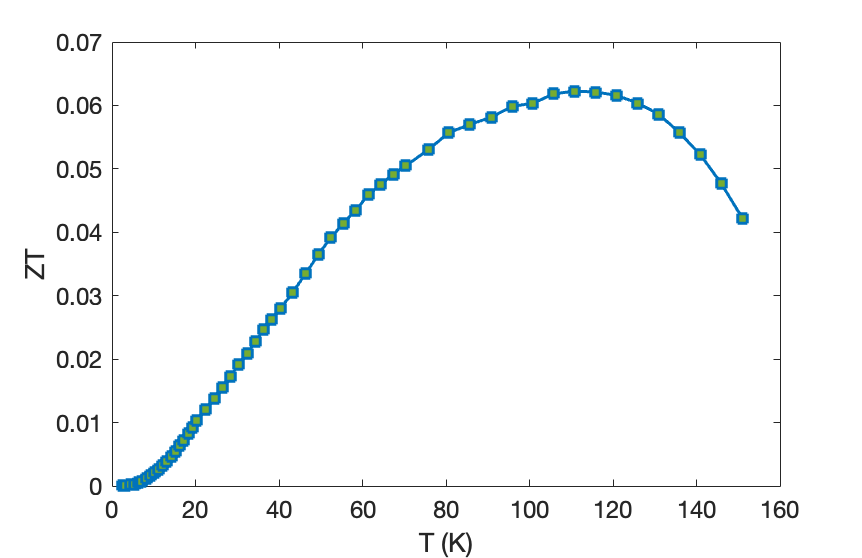}
    \caption{Thermoelectric figure of merit $\mathrm{ZT}$ for Ru$_2$Sn$_{3}\text{-}\#2$.}
    \label{fig:6}
\end{figure}

For completeness, before concluding, we estimate the thermoelectric figure of merit,
\begin{equation}
\mathrm{ZT} \equiv \frac{S^{2}\sigma}{\kappa} T,
\end{equation}
for Ru$_2$Sn$_3$\#2, as shown in Fig.~\ref{fig:6}. Both the magnitude and temperature dependence of $\mathrm{ZT}$ are consistent with those reported in Ref.~\cite{Kawasoko_2014}. We find that $\mathrm{ZT}$ reaches a maximum value of approximately $0.06$ at around $120$ K.

The relatively low $\mathrm{ZT}$ arises from the modest Seebeck coefficient and electrical conductivity, despite the comparatively low thermal conductivity. Achieving a high thermoelectric figure of merit requires the simultaneous optimization of electrical and thermal transport. In Ru$_2$Sn$_3$, although the lattice thermal conductivity is intrinsically low, its metallic character results in a sizable electronic contribution to the thermal conductivity, which limits overall performance.

From this perspective, Ru$_2$Sn$_3$ can be viewed as a parent compound: its low lattice thermal conductivity can be preserved while tuning the electronic properties to enhance efficiency. Strategies such as chemical doping, hot pressing, and other microstructural engineering approaches provide promising routes toward this optimization.

\section*{Discussion} 
Combining experimental measurements of the heat capacity, thermal conductivity, electric resistivity, and thermoelectric transport coefficients, we have reported the emergence of glassy thermodynamic and
transport properties in NCL compounds, whose crystalline nature is confirmed by means of structural characterization. Exploiting DFT and AIMD simulations, our analysis indicates that the anomalous glassy behavior of these crystalline materials is driven by low-lying optical phonons and their avoided crossing with the acoustic mode of highest group velocity. This interaction strongly distorts the acoustic dispersion from its ideal linear form, further contributing to the observed excess.

Importantly, the striking similarities with structural glasses extend beyond thermal transport and appear also in the electric and thermoelectric transport properties. It is interesting to notice that a linear magnetoresistance has been reported in Ru$_2$Sn$_{3}$ \cite{PhysRevB.101.205123} similarly to many strongly correlated electronic systems, such as cuprates, pnictides, and moire materials, where it is rationalized with the emergence of glassy orders \cite{doi:10.1073/pnas.2405720121}.

In conclusion, this large set of experimental and theoretical observations hints towards the emergence of surprising connections between crystals, glasses, and strange metals. In particular, there is increasing evidence that glassy physics can emerge without significant structural disorder and that strange metallic physics can arise without strong electronic correlations. In this direction, NCL crystals represent a perfect platform to reveal the microscopic origin of these exotic phenomena.

Finally, we anticipate that the peculiar effect of overdamped phonons on electrical transport might extend to a broader class of systems. In this context, it is worth noting that Ref.~\cite{boschi2025phason} recently reported phason-driven transport in moiré graphene, where the phason acts as an overdamped bosonic mode providing an efficient scattering channel, closely analogous to the low-lying optical modes discussed here. More generally, similar mechanisms may be relevant in charge-density-wave metals, strange metals, quantum paraelectric metals, and metallic glasses, where strong electron–phonon coupling, disorder, or proximity to electronic or structural instabilities can render phonons overdamped. Recent preprints \cite{gholap2026non,guo2026can} further support this perspective by highlighting the role of overdamped phonons in non-Fermi-liquid behavior near criticality. Exploring these connections represents a promising direction for future work.

%%%%%%%%%%%%%%%%%%%%%%%%
\section*{Methods}

\subsubsection*{Synthesis}
High-quality Ru$_2$Sn$_{3}$ single crystals were grown by a Bi flux method. Polycrystalline Ga-doped samples were produced by solid-state synthesis. See the SM for more details.

\subsubsection*{Experiments}
Resistance, heat capacity, thermoelectric and thermal transport measurements were
performed on a Physical Properties Measurement System (Quantum Design, PPMS). Site disorder and lattice parameter information for Ru$_2$Sn$_3$ samples were obtained via single crystal X-Ray Diffraction. Lattice parameters for Ga-doped samples were obtained via powder X-Ray Diffraction. See the SM for more details.

\subsubsection*{First-principles calculation}
First-principles calculation were performed on orthorhombic Ru$_2$Sn$_3$ using DFT and AIMD. See the SM for more details.

\section*{Data availability}
The datasets generated and analyzed during the current study are available upon reasonable request by contacting the corresponding authors. 

\section*{Code availability}
The codes that support the findings of this study is available upon reasonable request by contacting the corresponding authors. 

\section*{Acknowledgements}
SVM acknowledges the support of the Churchill Scholarship. WD and MB acknowledge the support of the Shanghai Municipal Science and Technology Major Project (Grant No.2019SHZDZX01). MB acknowledges the support of the sponsorship from the Yangyang Development Fund. XY and HX acknowledge the support of National Natural Science Foundation of China (Grant No. 12274283) and Science and Technology Commission of Shanghai Municipality (Grant No. 24LZ1401000). BZ and JM acknowledges the support of the National Key Research and Development Program of China (No. 2022YFA1402702). This work is supported by the National Key R$\&$D Program of China (Grant Nos. 2024YFA1409002, 2019YFA0308602), the National Natural Science Foundation of China (Grant No. 12074425), and the Fundamental Research Funds for the Central Universities, and the Research Funds of Renmin University of China (No. 23XNKJ22). We thank Dr. Meiyan Cui in Fujian Institute of Research on the Structure of Matter, Chinese Academy of Sciences, for the single-crystal XRD measurement and detailed refinement.

\section*{Author contributions}
MB, FMG, HX, JM supervised this project.
SVM synthesized and measured the heat capacity of sample $\#$1 and Ga-doped materials. SVM also measured resistivity of sample $\#$1 and conducted the DFT calculations. WD did the theoretical analysis with the help of MB and SVM. XY performed the measurement of thermal conductivity, the thermal electricity, and also resistivity of sample $\#$2. BZ measured heat capacity of sample $\#$2. EDN performed the structural characterization of sample $\#$1. JL and TX synthesized and characterized sample $\#$2. ZZ performed the single-crystal XRD of sample $\#$2. XC synthesized and characterized sample of Mn$_4$Si$_7$. YW, CS and JY performed the DFT and AIMD computations. All authors contributed to the writing of the manuscript and the interpretation of the results.

\section*{Competing interests}
The authors declare no competing interests.

%%%%%%%%%%%%%%%%%%%%%%%%%%%%%%%%%%%%%%%% 
\newpage
\appendix 
\clearpage
\renewcommand\thefigure{S\arabic{figure}}    
\renewcommand\thetable{S\arabic{table}}  
\setcounter{figure}{0} 
\renewcommand{\theequation}{S\arabic{equation}}
\setcounter{equation}{0}
\renewcommand{\thesubsection}{SM\arabic{subsection}.}

\begin{center}
   {\huge \textbf{Supplementary Material}}
\end{center}

%\tableofcontents

\subsection{Synthesis}
\begin{figure}[h]
    \centering
    \includegraphics[height=0.4\linewidth]{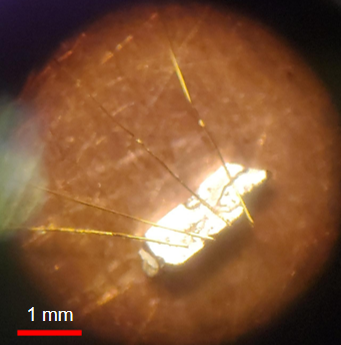}  \qquad   \includegraphics[height=0.4\linewidth]{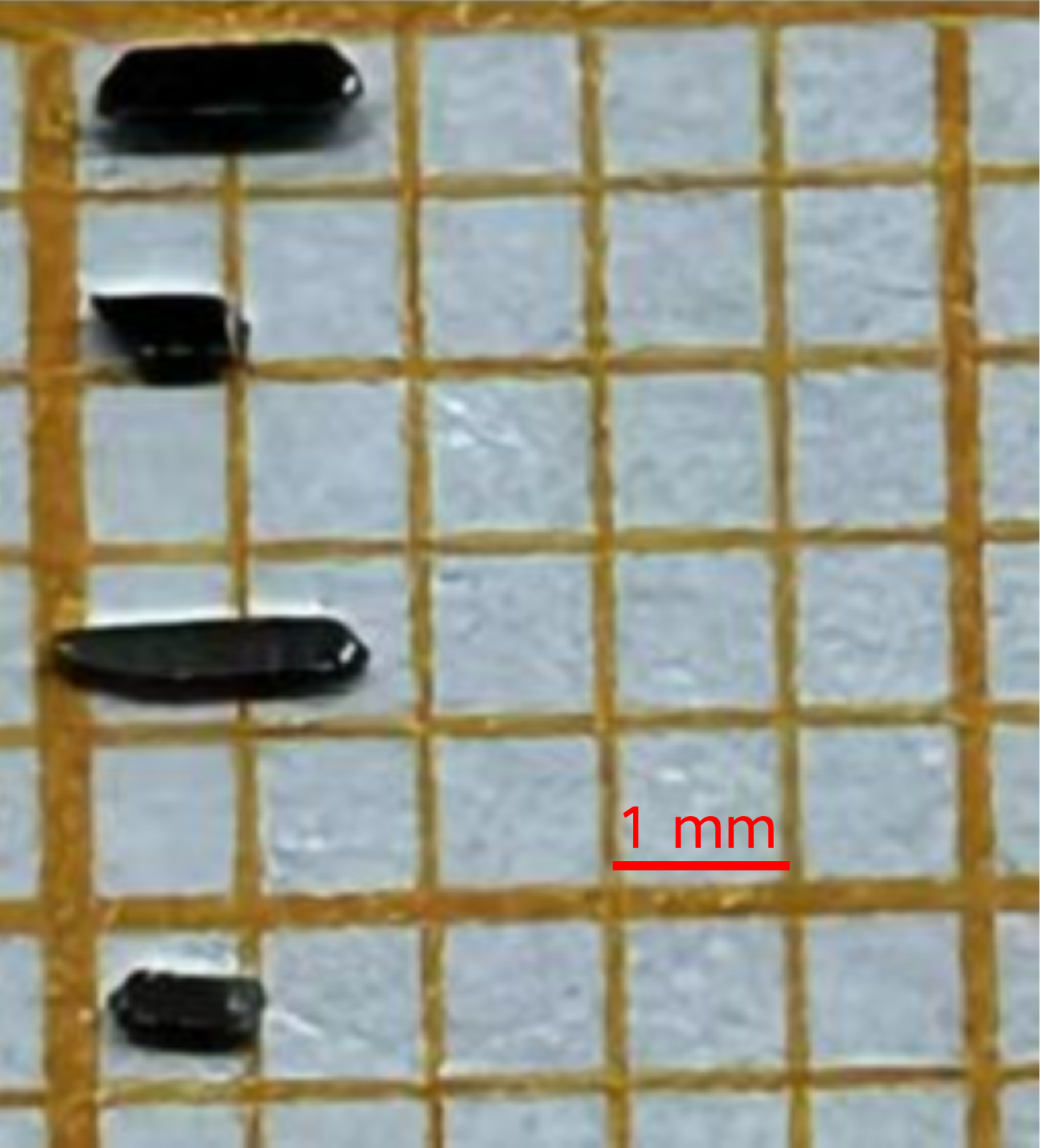}
    \caption{\textbf{Left: }Picture of Ru$_2$Sn$_{3}$ sample $\#1$, with gold transport wires attached. \textbf{Right: } Picture of Ru$_2$Sn$_{3}$ sample $\#2$.}
    \label{fig:sample}
\end{figure}

\textit{Growth Method Ru$_2$Sn$_{3}\text{-}\#1$} -- Single crystals of Ru$_2$Sn$_3$ (see Fig. \ref{fig:sample} for a picture of the sample) were synthesized by a Bi flux method as in Shiomi and Saitoh (2017) \cite{10.1063/1.4978773}.  Pure elemental Ru powder (99.95\%, Alfa Aesar), Sn pellets (99.99\%, Alfa Aesar) and Bi ingots (99.99\%, Alfa Aesar) were mixed in a molar ratio of Ru : Sn : Bi = 2 : 3 : 67 and loaded into an alumina crucible. Subsequently, the crucible was sealed into a quartz tube with a pressure less than $10^{-4}$ Pa. The tube was then put into a furnace, which was heated to 800 \textdegree C at 2 \textdegree C per minute and then kept there for 12 hours. The tube was then cooled to 100 \textdegree C over 70 hours. Single crystals were obtained by soaking the product in dilute HNO$_3$ acid.

\textit{Growth Method Ru$_2$Sn$_{3}\text{-}\#2$} --  High-quality Ru$_2$Sn$_{3}$ single crystals were grown by a Bi flux method. See Fig. \ref{fig:sample} for a picture of the sample. Pure elements of Ru powder, Sn powder and Bi granular were mixed in a molar ratio of Ru : Sn : Bi = 2 : 3 : 67 and loaded into a corundum crucible. Subsequently, the corundum crucible was sealed into a quartz tube with a pressure less than $10^{-4}$ Pa. The tube was then put into a furnace, which was heated to 1100 \textdegree C over 15 h and held at 1100 \textdegree C for 20 h before the temperature was decreased to 500 \textdegree C at 3 \textdegree C/h. Eventually, the single crystals with metallic luster were obtained by centrifugation at 500 \textdegree C to remove excess flux. Energy dispersive X-ray spectroscopy (EDX, Oxford X-Max 50) was performed to determine the chemical composition, which is consistent with the composition of Ru$_2$Sn$_{3}$ within instrumental error. The single crystal X-ray diffraction (XRD) pattern was carried out by a Bruker D8 Advance X-ray diffractometer using Cu K$_\alpha$ radiation.

\textit{Growth Method polycrystalline Ga-doped Ru$_2$Sn$_3$} -- Polycrystalline Ga-doped Ru$_2$Sn$_3$ was synthesized as in Lu et al. \cite{lu02}. Ru powder (99.95\%, Alfa Aesar), Sn pellets (99.99\%, Alfa Aesar), and Ga ingots (99.99 \%, Alfa Aesar) were ground together before being placed in an alumina crucible and sealed in a quartz tube. The samples were heated to 950 \textdegree C in a furnace at a rate of 2 \textdegree C per minute. They were baked for 100 hours and then cooled at a rate of 0.5 \textdegree C to room temperature. The lattice parameters measured by XRD are shown in Table \ref{table:PXRD}.

\begin{table*}[b]
\caption{Lattice parameters for Ga-doped synthesized compounds measured by XRD.}
\label{table:PXRD}
\centering
\begin{tabular}{@{}lcccccc@{}}
\hline \\
  Sample & a (\AA) &   c (\AA)  \vspace{0.15cm}\\ 
 
 \hline \\ 
RuGa$_{0.081}$Sn$_{1.438}$  & 6.180 & 94.060 \\ \\
RuGa$_{0.24}$Sn$_{1.319}$ & 6.159 & 123.330 \\
 \hline
\end{tabular}%
\end{table*}

\subsection{Structural characterization and absence of notable disorder}

Single crystal XRD measurements were performed on Ru$_2$Sn$_3$ samples to gain precise information on the crystals' structure and stoichiometry.

\textit{Measurements for sample $\#1$} -- XRD was performed on a Xcalibur Gemini ultra four-circle diffractometer equipped with an AtlasS2 detector and a graphite monochromator, using Mo-K$\alpha$ radiation ($\lambda$=0.71073 $\AA$). The integration of the data were performed using CrysAlisPro software (Rigaku Oxford Diffraction, 2022) and the absorption correction was performed using the SCALE3 ABSPACK scaling algorithm (empirical absorption correction using spherical harmonics) implemented in CrysAlisPro. Structural resolution and refinement were done using Jana2020 software \cite{jana2020} in the P$\overline{4}$c2 space group (116) leading to an agreement factor R = 0.0258, reflecting the good quality of the structure solution. Occupancies were refined and converged to a sum formula of Ru$_{1.989}$Sn$_{2.984}$. The ADP (atomic displacement parameters) from the refined structure do not present any singularity or anisotropic shape suggestive of structural disorder (CCDC deposition number of the structure: 2452973).

\textit{Measurements for sample $\#2$} -- XRD data were collected by a XtaLAB Synergy R, HyPix four-circle diffractometer with a graphite-monochromated Mo-K$\alpha$ radiation ($\lambda$ = 0.71073 $\AA$). The structure was conducted by software Olex2 ~\cite{Dolomanov} using the Rietveld method, and the final information was checked by PLATON~\cite{Spek}. Lattice parameters are reported in Table \ref{tab:crystal_structure}. The R factor was smaller than $0.03$, demonstrating good data quality and the reliability of the refinement results. The crystal structure, refined in the space group P$\overline{4}$c2, is presented in Table~\ref{tab:crystal_structure} and the detailed bond length estimates are reported in Table~\ref{tab:bond_lengths}. Occupancies were refined and converged to a sum formula of Ru$_{1.998}$Sn$_{2.999}$. 
\\
The data points in Fig. \ref{fig:ref} confirm long-range lattice order for both samples. Based on the lattice symmetry, the Sn atoms occupying the Wyckoff $\textit{4e}$ and $\textit{8j}$ sites were designated as Sn1 and Sn2, respectively, while the Ru atoms at the $\textit{2d}$, $\textit{4i}$, and $\textit{2a}$ positions were labeled as Ru1, Ru2, and Ru3, respectively. Most of atomic bond lengths show very minimal variations across the two samples. 

\textit{Measurements for Ga-doped samples} -- XRD data were collected on ground up powders, in a Bruker D8 diffractometer. Data were refined in TOPAS using the Pawley method to obtain lattice parameters.

The single-crystal X-ray diffraction analysis and the corresponding structural refinements confirm the long-range crystalline order of both Ru$_2$Sn$_3$ samples and rule out the presence of any detectable amorphous phase.

\begin{table}
\caption{Crystal structure of Ru$_{2}$Sn$_{3}\text{-}\#$1 and Ru$_{2}$Sn$_{3}\text{-}\#$2 determined by single-crystal XRD. Listed are the measurement temperature ($T$), lattice constants ($a$, $c$), space group, and atomic occupancy (Occ) for Ru and Sn.}
\centering
\begin{tabular}{@{}lcccccc@{}}
\hline \\
Sample & $T$ (K) & $a$ (\AA) & $c$ (\AA) & Space group & Occ(Ru) (\%) & Occ(Sn) (\%) \vspace{0.15cm}\\ 
 \hline \\ 
Ru$_2$Sn$_{3}$-\#1 & 299 & 6.1695(4) & 9.9201(7) & P$\overline{4}$c2 & 99.45 & 99.47 \\ \vspace{0.01cm} \\
Ru$_2$Sn$_{3}$-\#2 & 292 & 6.1791(1) & 9.9271(5) & P$\overline{4}$c2 & 99.90 & 99.97 \\
 \hline
\end{tabular}
\label{tab:crystal_structure}
\end{table}

\begin{table}
\caption{Detailed bond lengths (\AA) of Ru$_{2}$Sn$_{3}$-\#1 and Ru$_{2}$Sn$_{3}$-\#2 determined by single-crystal XRD.}
\centering
\begin{tabular}{ccc}
  \hline \\
Bond Pair & Ru$_{2}$Sn$_{3}$-\#1 & Ru$_{2}$Sn$_{3}$-\#2 \vspace{0.15cm} \\
\hline \\
Sn1-Sn1    & 3.1254(10) & 3.1230(2)  \\\\
Sn1-Sn2    & 3.1429(11) & 3.1476(11) \\\\
Sn1-Ru1    & 2.9313(5)  & 2.9321(6)  \\\\
Sn1-Ru2    & 2.5857(10) & 2.5899(8)  \\\\
Sn1-Ru3    & 2.7998(7)  & 2.8079(11) \\\\
Sn2-Sn2    & 3.2416(12) & 3.2433(19) \\\\
Sn2-Ru1    & 2.6585(8)  & 2.6624(8)  \\\\
Sn2-Ru2    & 2.7533(8)  & 2.6690(17) \\\\
Sn2-Ru3    & 2.5642(9)  & 2.5670(11) \\
  \hline
\end{tabular}
\label{tab:bond_lengths}
\end{table}
\begin{figure}[h]
    \centering
\includegraphics[width=\linewidth]{fig_reply_5.png}
    \caption{The measured structural factors in arbitrary units, $F^2_{\text{mea}}$, are plotted versus the calculated ones, $F^2_{\text{cal}}$, for the final refinement of \textbf{(a)} Ru$_2$Sn$_{3}\text{-}\#1$ at 299 K and \textbf{(b)} Ru$_2$Sn$_{3}\text{-}\#2$ at 292 K. The red solid line shows the good agreement between the measurement and calculations.}
    \label{fig:ref}
\end{figure}

Additional details on the crystal structures, including unit cell parameters, atomic coordinates, and symmetry information, are provided in the CIF file available as Supplementary Material. 
\subsection{Experimental details}
\textit{Experimental measurements for sample $\#1$} --  Measurements of four-terminal AC electrical transport and heat capacity were performed on a Physical Properties Measurement System (Quantum Design). Powder X-ray diffraction measurements were performed at room temperature on a Xcalibur Gemini ultra four-circle diffractometer. Structural refinements were performed using TOPAS software.

\textit{Experimental measurements for sample $\#2$} -- Heat capacity measurement for Ru$_2$Sn$_{3}\text{-}\#2$ in 2 - 100 K was performed by thermal relaxation method using the physical property measurement system (Quantum Design, PPMS). Thermoelectric and thermal transport measurements are performed on the same sample in a Quantum Design PPMS, using the same contacts made of silver paste Dupont 4929N. For thermal measurements, a constant heat current $Q$ applied by a chip resistance was sent in the basal plane of the single crystal, generating a longitudinal temperature gradient $\nabla_x T=\frac{T_1-T_2}{l}$, where $l$ is the spacing between the two temperature measurement points along the x-axis. The thermal conductivity along the x-axis can be given by $\kappa_{x x}=\frac{Q}{\nabla_x T \cdot w t}$, where $w$ is the width of the sample (alone y-axis), $t$ is the thickness of the sample. The thermoelectric coefficients are given by measuring the electric potential difference generated by the longitudinal temperature gradient $\nabla_x T$. The Seebeck coefficient (longitudinal thermoelectric coefficient) $S$ can be given by $S=\frac{U_x}{\nabla_x T \cdot d_x}$, where $U_x$ is the voltage difference between two points parallel to the x-axis and $d_x$ is the distance between the two points. Like the Seebeck coefficient, the Nernst coefficient (transverse thermoelectric coefficient) $N$ is given by $N=\frac{U_y}{\nabla_x T \cdot d_y}$ with applied magnetic field along the z-axis, where $U_y$ is the voltage difference between two points perpendicular to the x-axis and $d_y$ is the distance between the two points.

\begin{figure}
    \centering
    \includegraphics[width=\linewidth]{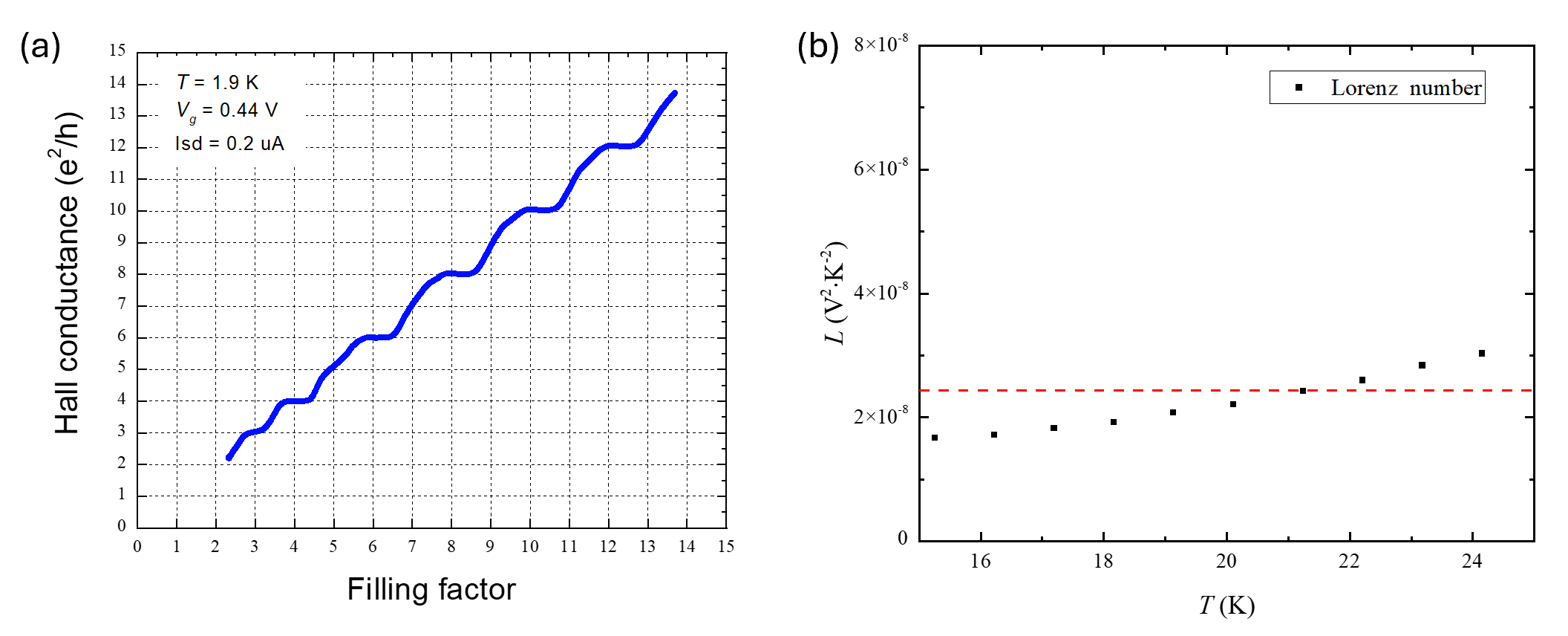}
    \caption{\textbf{(a)} Resistivity calibration based on the quantum Hall effect in a two-dimensional electron gas (2DEG).
The sample consists of a GaAs/AlGaAs heterostructure hosting a high-mobility 2DEG with electron density
$n_{2D} = 5.7 \times 10^{11}\,\mathrm{cm^{-2}}$ and mobility
$\mu = 1.89 \times 10^{5}\,\mathrm{cm^{2}/Vs}$, measured at $77\,\mathrm{K}$.
A transparent NiCr top gate is used to tune the carrier density.
\textbf{(b)} Thermal conductivity calibration based on the Wiedemann--Franz law.
The Fermi-liquid state in $2M$-WS$_2$ is used to calibrate the thermal conductivity measurements at low temperatures; see \cite{Yang2023} for details.
}
    \label{fig:calib}
\end{figure}

The electrical resistivity and thermal transport data were not measured using the standard thermal transport option (TTO) of the PPMS, but with a home-built sample puck based on standard resistivity/Hall puck configurations. The electrical resistivity was calibrated using the quantum Hall plateau, while the thermal conductivity was benchmarked against the Wiedemann--Franz law in the low-temperature Fermi-liquid regime, as shown in Fig.~\ref{fig:calib}. The uncertainty in the resistivity is estimated to be approximately 2\%, mainly due to the sample geometry. For the thermal conductivity, the main sources of uncertainty include heat losses, contact thermal resistance, thermometer calibration, and sample geometry, leading to an estimated upper bound of approximately 20\%.

All transport measurements were performed in an in-plane geometry, with the applied current and thermal gradient lying in the basal plane of the single crystal.

\subsection{First-principles calculations}

First-principles simulations were performed using the Vienna \emph{ab initio} simulation package (VASP) \cite{PhysRevB.54.11169,KRESSE199615}, employing the projector-augmented wave (PAW) method \cite{PhysRevB.59.1758}. The exchange–correlation functional was treated within the Perdew–Burke–Ernzerhof (PBE) approximation \cite{PhysRevLett.77.3865}. The orthorhombic Ru$_2$Sn$_3$ structure was fully relaxed until the total energy converged below $5\times10^{-8}$ eV.

To investigate finite-temperature effects on phonons, AIMD simulations were carried out. The system was first equilibrated in the isothermal–isobaric (NPT) ensemble at 100 K for 5 ps. Subsequently, simulations were performed in the canonical (NVT) ensemble using a Nosé–Hoover thermostat at 100 K for 30 ps, with a time step of 2 fs. A $2\times2\times3$ supercell containing 480 atoms was employed.

The second-order interatomic force constants (IFCs) were extracted using the \texttt{hiPhive} package \cite{https://doi.org/10.1002/adts.201800184} and compared with the corresponding $T=0$ K DFT results. Phonon spectra were calculated using the \texttt{Phonopy} package \cite{TOGO20151,doi:10.7566/JPSJ.92.012001}. The lattice thermal conductivity, as well as the phonon group velocity and scattering rates at 100 K, were computed using the \texttt{ShengBTE} package \cite{LI20141747}, based on second- and third-order IFCs derived from the AIMD simulations.

\subsection{Finite-temperature and anharmonic effects}

 \label{SM:AIMD}

\begin{figure}
    \centering
    \includegraphics[width=0.9\linewidth]{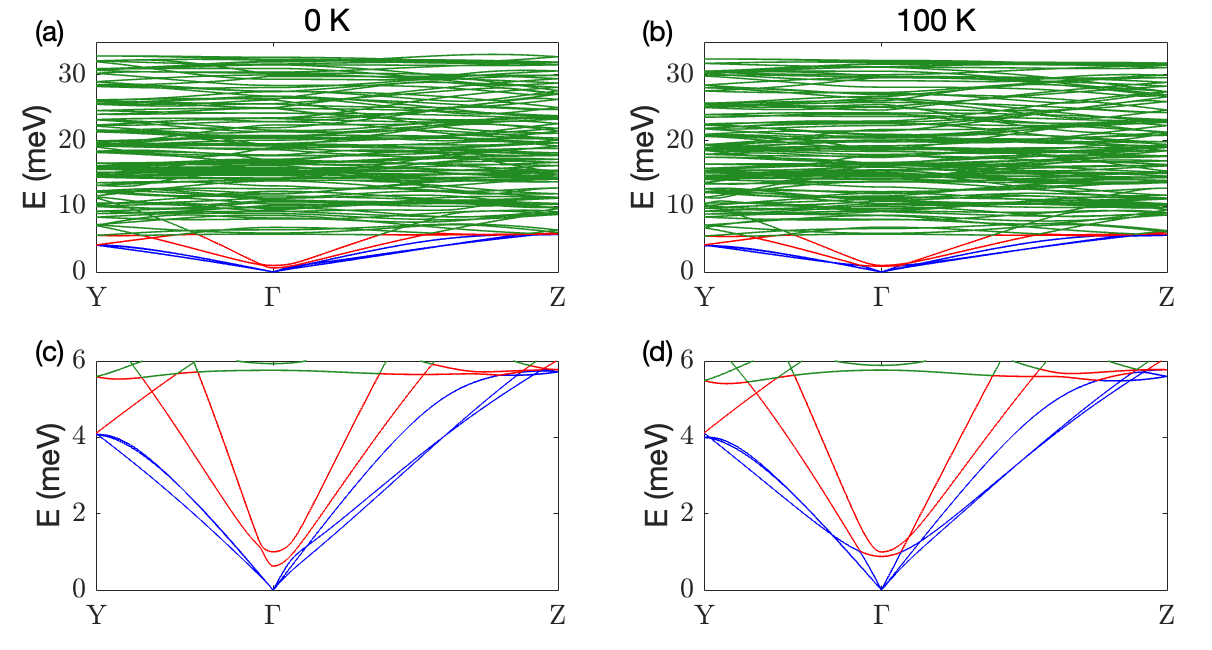}
    \caption{Vibrational spectra of Ru$_2$Sn$_3$ at different temperatures.
\textbf{(a)} Full vibrational spectrum at $T=0$ K obtained from DFT.
\textbf{(b)} Full vibrational spectrum at $T=100$ K obtained from AIMD.
\textbf{(c)} Enlarged view of the low-energy region at $T=0$ K.
\textbf{(d)} Enlarged view of the low-energy region at $T=100$ K.}
    \label{0K100K}
\end{figure}

To assess finite-temperature effects, we also performed AIMD simulations at $T=100$ K.
As shown in Fig.~\ref{0K100K}, the overall vibrational spectrum remains qualitatively similar to that at $T=0$ K, although the two lowest optical modes become less separated from the acoustic branches and start to hybridize with them at finite temperature. 

\begin{figure}
    \centering
    \includegraphics[width=1\linewidth]{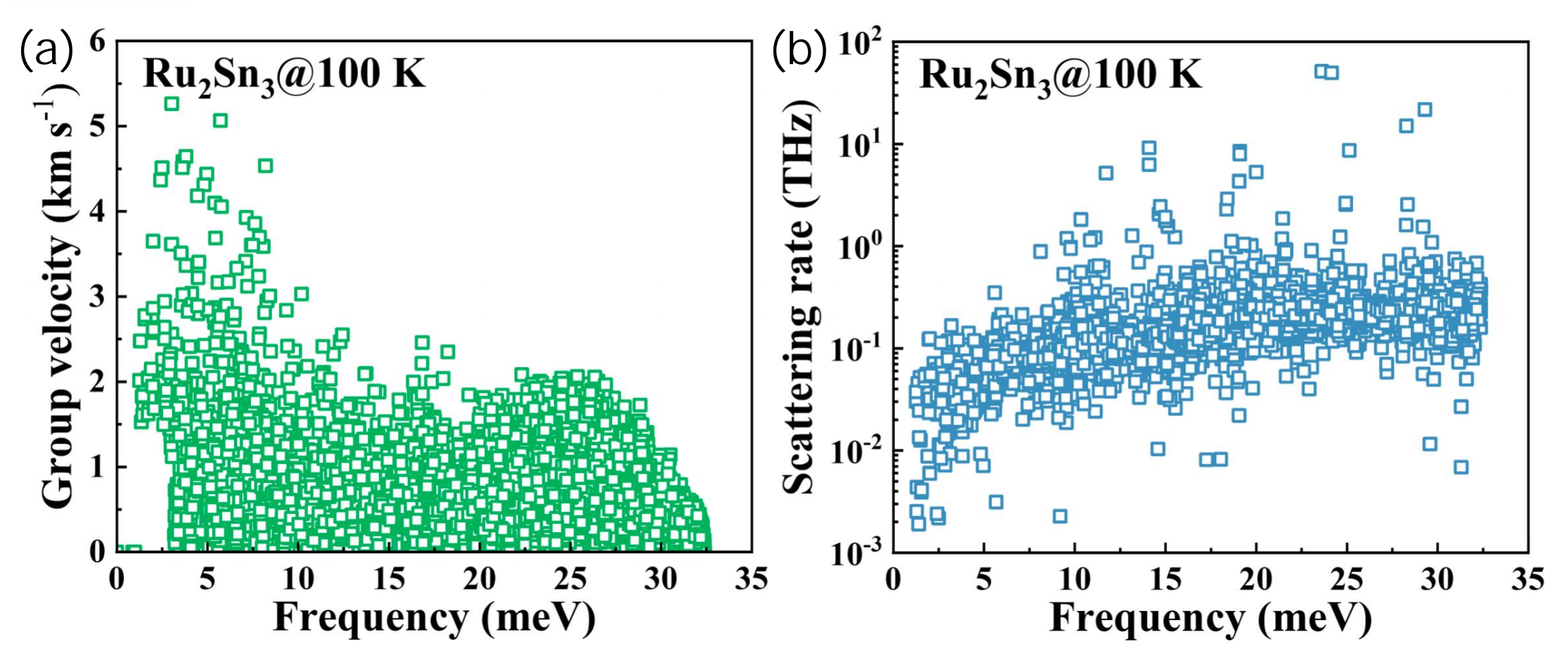}
    \caption{\textbf{(a)} Mode phonon group velocity and \textbf{(b)} scattering rate as the function of phonon frequency at 100 K.}
    \label{fig:group_velocity_scattering}
\end{figure}

Fig.~\ref{fig:group_velocity_scattering} reports the phonon group velocity and scattering rate as functions of frequency at 100 K. The data indicate that, while the highest group velocity reaches 5.27 km s$^{-1}$, the vast majority of phonon modes are confined below 3 km s$^{-1}$. This is especially true in the low-to-medium-frequency range, which provides the dominant contribution to thermal transport. 

When weighted by the scattering rate, the average phonon group velocity is found to be approximately 0.60 km s$^{-1}$. This value is markedly smaller than that of typical high-$\kappa_l$ systems, such as FeSi$_2$, where the average reaches 1.26 km s$^{-1}$ and the lattice thermal conductivity is 24.26 W m$^{-1}$ K$^{-1}$ at 300 K \cite{Zhao2017}.
\subsection{Low temperature analysis}

\begin{figure}[ht]
    \centering
\includegraphics[width=1\linewidth]{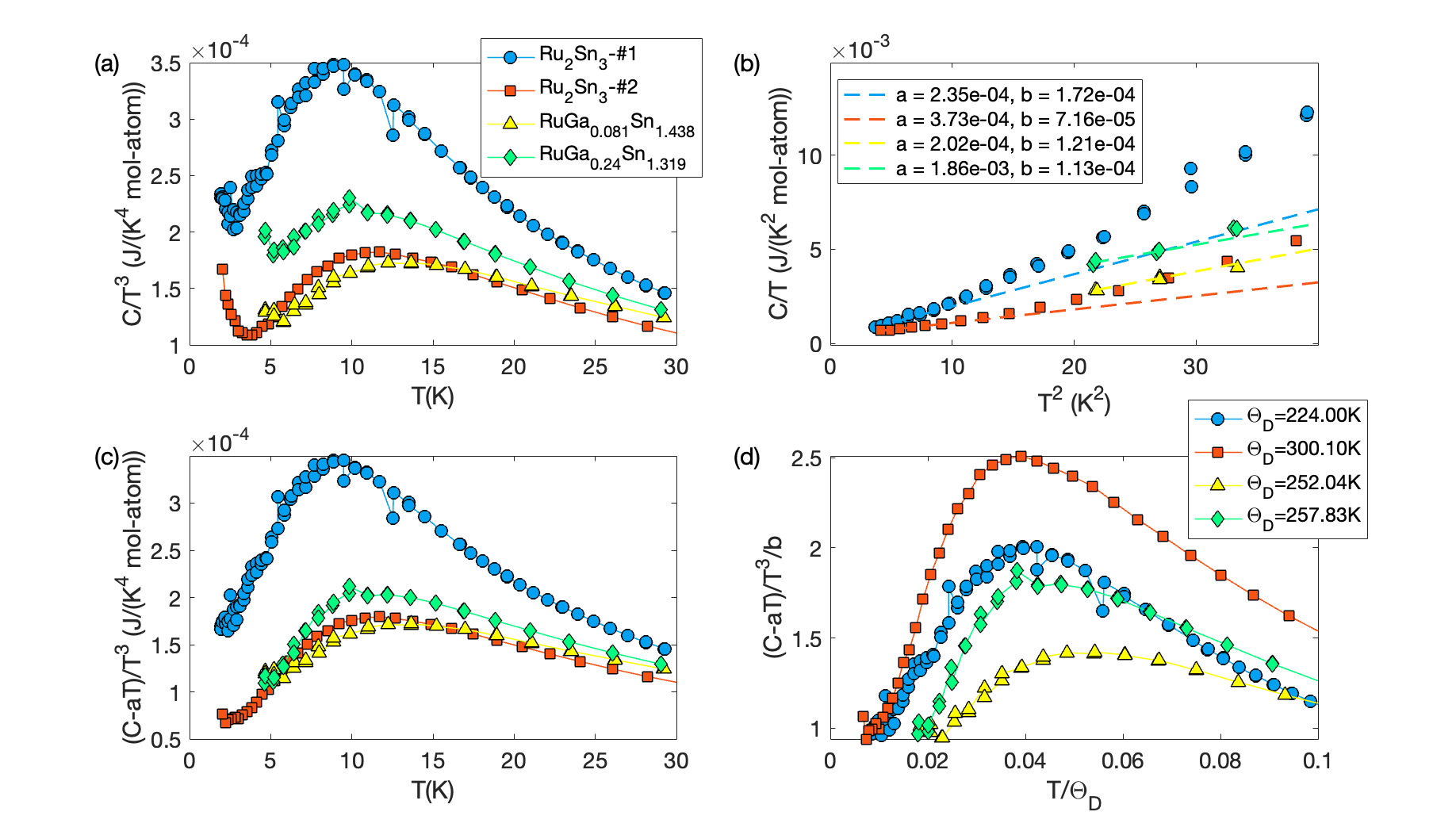}\caption{\textbf{(a)} Measured heat capacity $C(T)$ normalized by $T^3$
for several NCL materials. \textbf{(b)} $C(T)/T$ as a function of $T^2$ and numerical fits to the function $C(T)/T=a+bT^2$ (dashed lines). \textbf{(c)} Experimental data with electron contribution subtracted, $C(T)-aT$, and normalized by $T^3$. \textbf{(d)} $(C(T)-aT)/T^3$ normalized by $b$ as a function of $T/\Theta_D$.} 
    \label{fig:s4}
\end{figure}

\begin{figure}[h]
    \centering
    \includegraphics[width=0.8\linewidth]{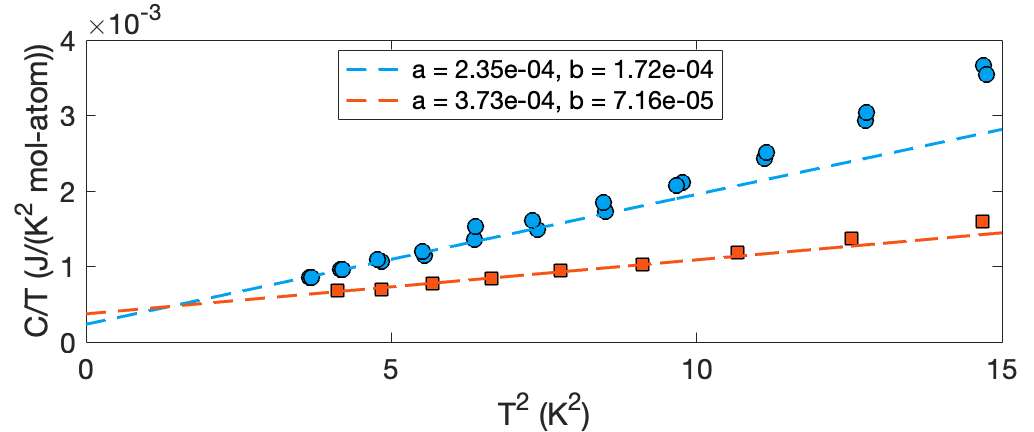}
    \caption{A zoom of $C(T)/T$ as a function of $T^{2}$ for Ru$_2$Sn$_{3}\text{-}\#1$ (blue) and Ru$_2$Sn$_{3}\text{-}\#2$ (red). The dashed lines show numerical fits to the function $C(T)/T = a + b T^{2}$, where the first term is the electronic Sommerfeld contribution and the second one the phononic Debye term.}
    \label{fig:C/T}
\end{figure} 

\begin{center}
\begin{table}[h]
    \centering
    \begin{tabular}{|c|c|c|}
        \hline
        Material & $a$ (J/K$^2$ mol-atom) & $b$ (J/K$^4$ mol-atom) \\ 
        \hline
        Ru$_2$Sn$_{3}\text{-}\#1$ &2.35$\times$10$^{-4}$&1.72$\times$10$^{-4}$\\
        \hline
        Ru$_2$Sn$_{3}\text{-}\#2$ &3.73$\times$10$^{-4}$&7.16$\times$10$^{-5}$\\
        \hline
RuGa$_{0.081}$Sn$_{1.438}$ &2.02$\times$10$^{-4}$&1.21$\times$10$^{-4}$\\
        \hline
        RuGa$_{0.24}$Sn$_{1.319}$ &1.86$\times$10$^{-3}$&1.13$\times$10$^{-4}$\\
        \hline
    \end{tabular}
    \caption{Power-law fitting parameters using the low-temperature polynomial form $C(T)=a T+ b T^3$ and the experimental data at low temperature.}
    \label{fit}
    \end{table}
\end{center}
In Fig.~\ref{fig:s4}(a), we display  the raw experimental data of $C(T)/T^3$ as a function of $T$. In panel (b) of the same figure, we show a different representation of the data by plotting $C(T)/T$ as a function of $T^2$. The dashed lines correspond to the fits to the polynomial function $C(T)/T=a+bT^2$. The obtained values for the fitting parameters $a$ and $b$ are presented in Table \ref{fit}.

The experimental heat capacity with the subtracted electronic Sommerfeld contribution, $(C(T)-a T)/T^3$ is shown in Fig.~\ref{fig:s4}(c). The low-temperature data clearly confirm the presence of a Sommerfeld contribution linear in $T$, together with a phononic Debye term proportional to $T^{3}$. Finally, in panel (d) of the same figure, the same experimental data are represented by normalizing the temperature to the Debye temperature $\Theta_D$. The Debye temperature is calculated using $\Theta_D=(\frac{12\pi^4}{5}\frac{N_{mol}k_B}{b})^{1/3}$.
We notice that after normalizing the $x$-axes by the Debye temperature, the position of the peak for the two Ru$_2$Sn$_3$ samples is nearly the same. On the other hand, the density change encoded in $\theta_D$ is not enough to account for the energy shift of the peak in the Ga-doped samples. This is consistent with the results presented in Ref. \cite{Miyazaki}. 

In Fig.~\ref{fig:C/T}, we present a zoomed view of $C/T$ as a function of $T^2$ for undoped Ru$_2$Sn$_3$ crystals, highlighting the presence of a Sommerfeld contribution, consistent with their metallic character and electrical resistivity measurements.

\subsection{Extended analysis: electric resistivity}
 \begin{figure}[ht]
    \centering
\includegraphics[width=0.9\linewidth]{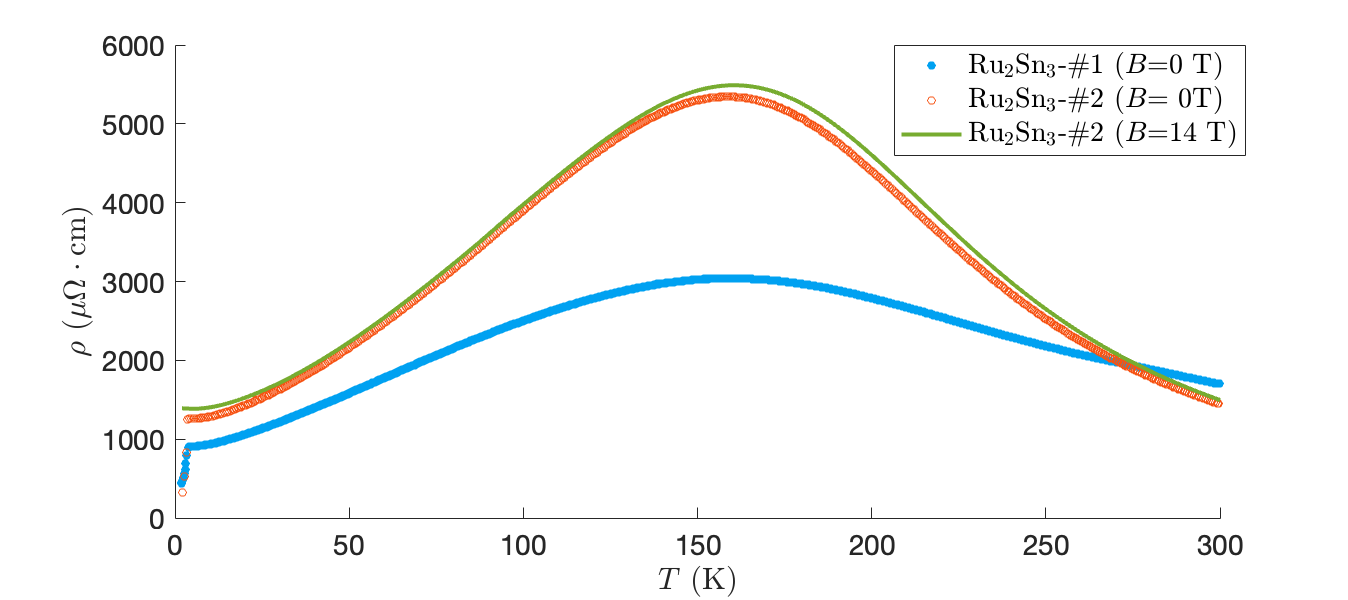}
\caption{Measured resistivity $\rho(T)$ for Ru$_2$Sn$_{3}\text{-}\#1$ up to 300 K at 0 T and Ru$_2$Sn$_{3}\text{-}\#2$ up to 300 K at 0 T and 13.9 T magnetic field.}
    \label{SM:rho1}
\end{figure}    

In Fig.~\ref{SM:rho1}, we show the measured electric resistivity $\rho(T)$ for Ru$_2$Sn$_{3}\text{-}\#2$ at zero magnetic field (orange) and $B=13.9$ T magnetic field (green). Two points are worth to be noticed. First, at zero magnetic field, the material exhibits a superconducting phase transition around $T_c \approx 3.7$ K that is signaled by a sudden drop in the electric resistivity. This transition has been attributed to Sn impurities, but this is still in dispute \cite{10.1063/1.4978773, PhysRevB.101.205123}. Moreover, both measurements show a peak at $\approx$ 160 K, which is consistent with a structural phase transition from orthorhombic at low T to tetragonal at high T and also reported in Refs.~\cite{10.1063/1.4978773,gibson_quasi_2014,PhysRevB.101.205123}. As discussed in the main text, this structural phase transition is responsible for one of the two low-lying optical modes in Ru$_2$Sn$_{3}$.

\begin{figure}[ht]
    \centering
\includegraphics[width=\linewidth]{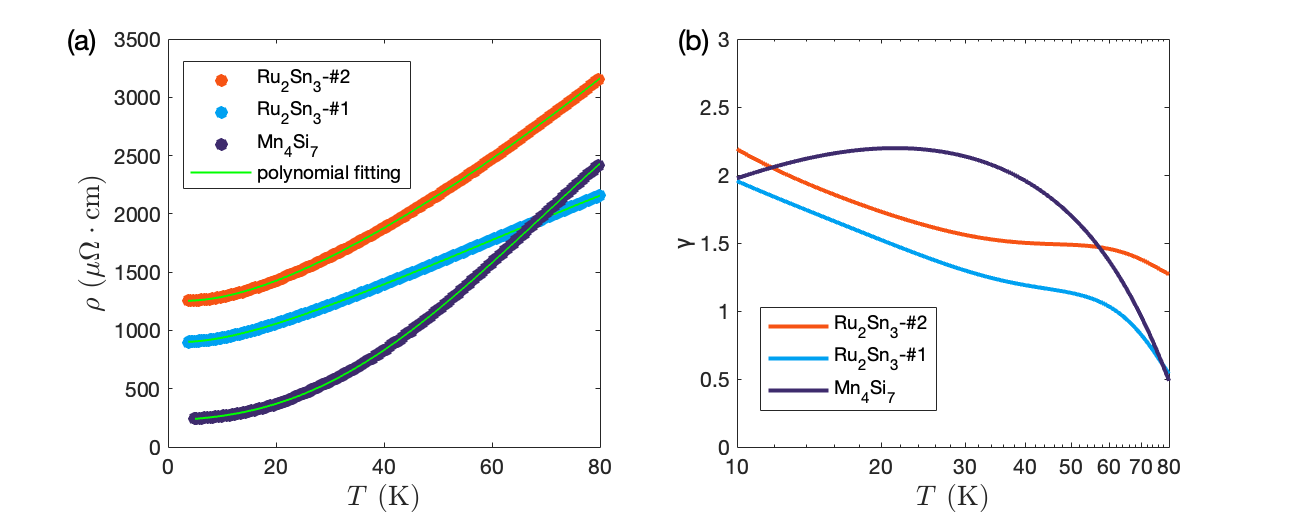}
\caption{\textbf{(a)} Measured resistivity $\rho(T)$ for Ru$_2$Sn$_{3}\text{-}\#1$, Ru$_2$Sn$_{3}\text{-}\#2$, and Mn$_4$Si$_7$ and polynomial fittings. The data for Mn$_4$Si$_7$ are taken with permission from Ref. \cite{Chen2015}. \textbf{(b)} Scaling parameter $\gamma$ calculated with polynomial fittings, see text for details (Eq.~\eqref{lalal}).}
    \label{SM:rho2}
\end{figure}    

In Fig.~\ref{SM:rho2}(a), we show the measured resistivity for Ru$_2$Sn$_{3}\text{-}\#1$, Ru$_2$Sn$_{3}\text{-}\#2$ and Mn$_4$Si$_7$. We show that all three NCL compounds show a clear $T^2$ to linear crossover and an extended and robust linear in $T$ regime. We use a polynomial fit to describe the experimental data and extract the phenomenological scaling parameter:
\begin{equation}\label{lalal}
\gamma=1+\frac{T \frac{d^2 \rho}{d T^2}}{\frac{d \rho}{d T}}.
\end{equation}
The trend of this scaling parameter as a function of temperature is shown in Fig.~\ref{SM:rho2}(b).

\subsection{Extended analysis: low-lying optical modes}
\begin{figure}[h]
    \centering
    \includegraphics[width=0.8\linewidth]{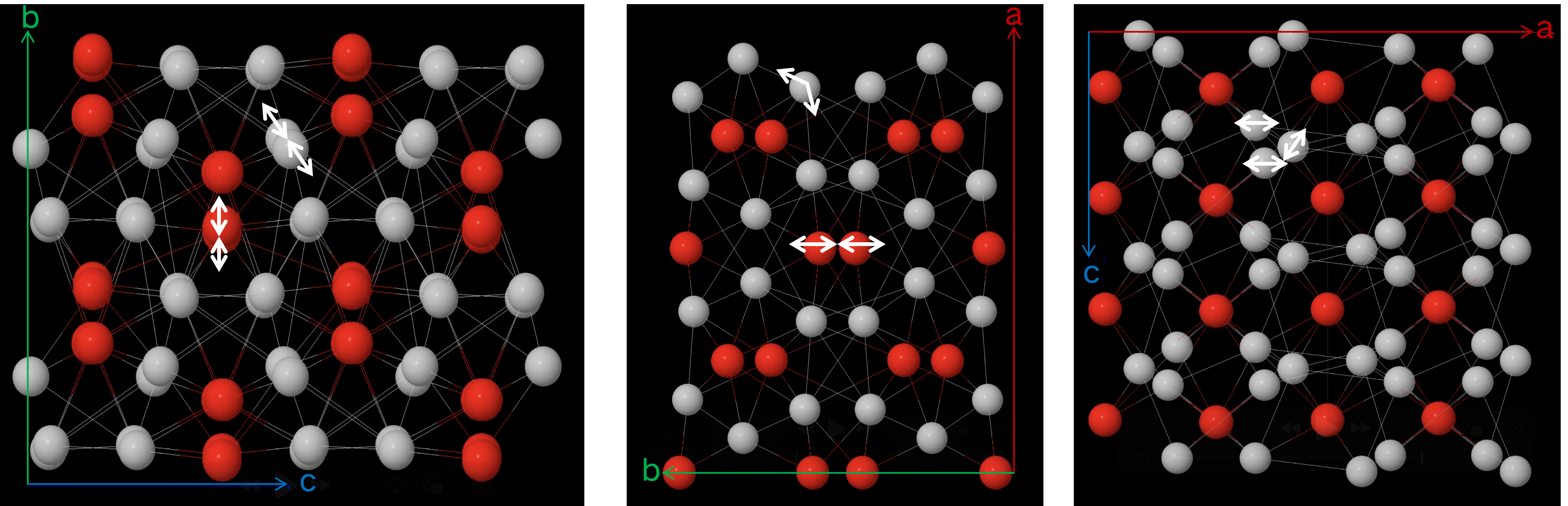}
    \caption{Snapshots from the animation of the Ru$_2$Sn$_{3}$ soft mode (referred to as 'second optical' in the main text), as calculated by DFT. Sn atoms (silver) oscillate between the positions they hold in the low-temperature orthorhombic phase and the high-temperature tetragonal phase. Ru atoms (red) are mostly static. The perspective is first along the a-axis, then along the c-axis, then along the b-axis. Animations were produced using Jmol software.}
    \label{fig:soft-mode-video}
\end{figure}

\begin{figure}[h]
    \centering
    \includegraphics[width=0.8\linewidth]{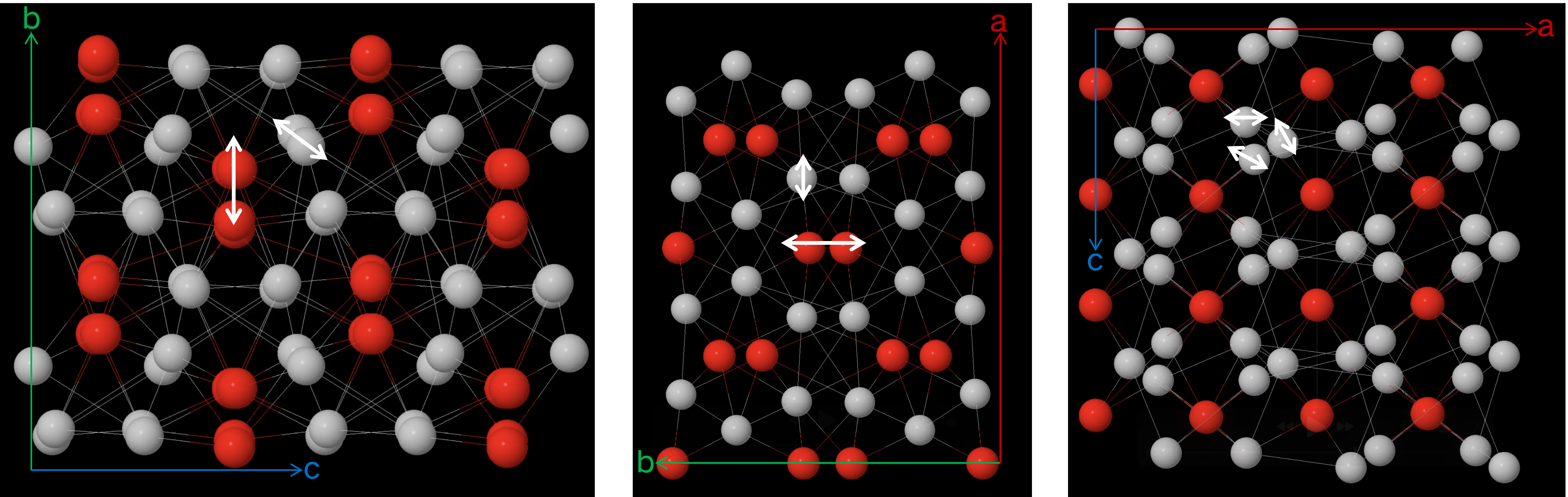}
    \caption{Snapshots from the animation of the Ru$_2$Sn$_{3}$ twisting mode (referred to as 'first optical' in the main text), as calculated by DFT. Sn atoms (silver) corkscrew through the lattice of mostly static Ru atoms (red). The perspective is first along the a-axis, then along the c-axis, then along the b-axis. Animations were produced using Jmol software.}
    \label{fig:twisting-mode-video}
\end{figure}

In Figs. \ref{fig:soft-mode-video}-\ref{fig:twisting-mode-video}, we provide three snapshots of the video animation for the two low-lying optical modes in Ru$_2$Sn$_{3}$, corresponding to the dispersion relations shown in red color in Fig. 2(c). Animation videos are provided as well in the attachment to this manuscript.

For the Ru$_2$Sn$_{3}$ soft mode (referred to as 'second optical' in the main text), Sn atoms (silver) oscillate between the positions they hold in the low-temperature orthorhombic phase and the high-temperature tetragonal phase. Ru atoms (red) are mostly static. On the other hand, for the Ru$_2$Sn$_{3}$ twisting mode (referred to as 'first optical' in the main text), Sn atoms (silver) corkscrew through the lattice of mostly static Ru atoms (red). This second mode is analogous to the twisting mode reported in Mn$_4$Si$_7$ in Ref. \cite{Chen2015}. Finally, the soft mode has Ag symmetry, while the twisting mode has B1u symmetry.

\subsection{Hall response}

\begin{figure}[t]
    \centering
    \includegraphics[width=\linewidth]{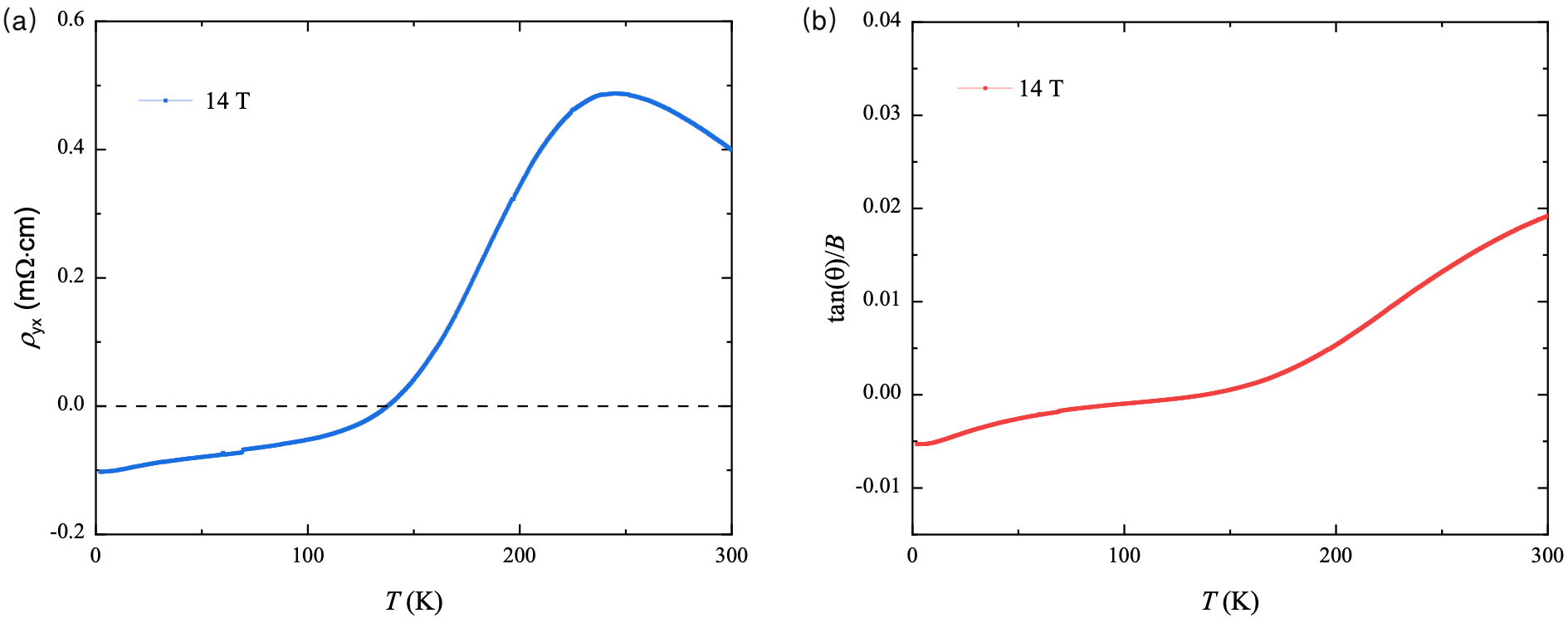}
    \caption{\textbf{(a)} In-plane Hall resistivity and \textbf{(b)} Hall angle as a function of temperature for Ru$_2$Sn$_3$-\#2 at 14~T.}
    \label{fig:Hall}
\end{figure}

To further assess whether the thermoelectric anomalies have a direct electronic origin, we also measured the Hall response of Ru$_2$Sn$_3$-\#2. As shown in Fig.~\ref{fig:Hall}, within the plateau-like temperature window identified from the thermoelectric response, both the Hall resistivity and the Hall angle show only weak temperature dependence and do not exhibit clear anomalous features. This observation is consistent with our interpretation that low-lying phonons play a key role in the anomalous thermoelectric response. If the anomalies were dominated by changes in the electronic band structure alone, one would in general expect a corresponding signature in the Hall response as well.

\subsection{Linear-linear representation of the thermal conductivity}

\begin{figure}[h]
    \centering
    \includegraphics[width=0.6\linewidth]{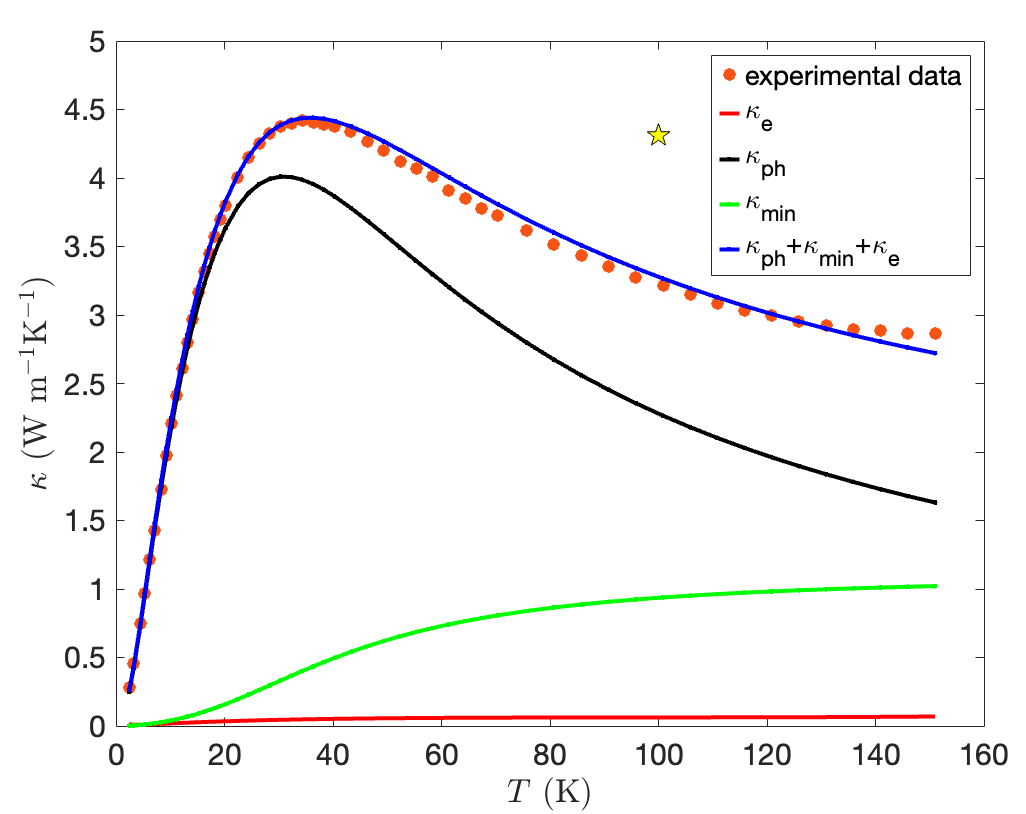}
    \caption{Linear-linear representation of the thermal conductivity of Ru$_2$Sn$_{3}\text{-}\#2$, including the electronic contribution $\kappa_{e}$ based on the Wiedemann--Franz law, the phononic contribution $\kappa_{\mathrm{ph}}$ described by the Debye--Peierls relaxation-time model, and the minimum thermal conductivity $\kappa_{\mathrm{min}}$ estimated using the Cahill--Pohl model. The yellow star indicates the calculated lattice thermal conductivity at 100 K using AIMD.}
    \label{fig:linear-k}
\end{figure}

To complement the logarithmic plots shown in the main text, we present in Fig.~\ref{fig:linear-k} the thermal conductivity of Ru$_2$Sn$_3$-\#2 on a linear–linear scale.

\subsection{Electronic band structure}

\begin{figure}[t]
    \centering
    \includegraphics[width=0.7\linewidth]{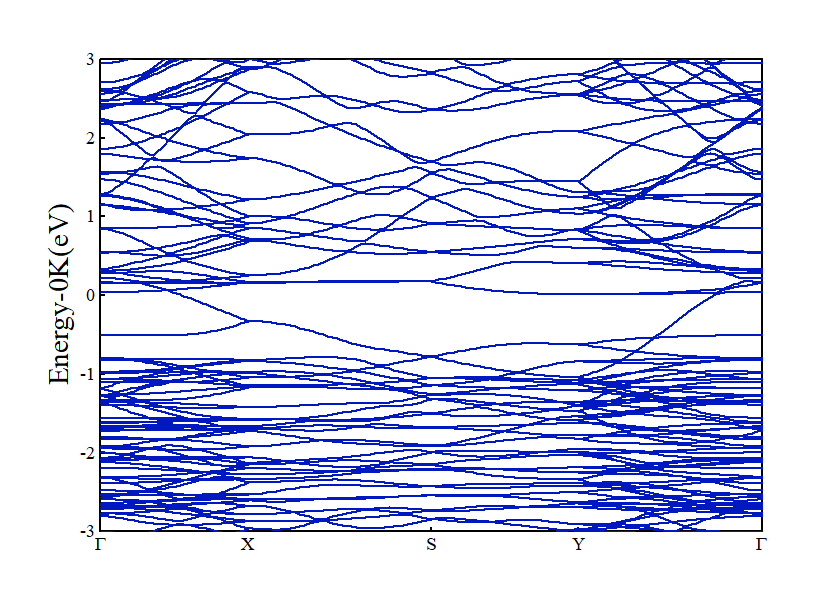}
    \caption{Electronic band structure of Ru$_2$Sn$_3$ obtained from DFT at $T=0$~K, confirming the metallic character of the compound.}
    \label{fig:eband}
\end{figure}

For completeness, we also performed standard DFT calculations of the electronic band structure. The result is shown in Fig.~\ref{fig:eband} and confirms the metallic nature of Ru$_2$Sn$_3$.

\end{document}